\documentclass[12pt,notoc]{JHEP3}

\usepackage{amsmath,amssymb,euscript,array,cite,mathrsfs}

\newcommand{\startappendix}{
\setcounter{section}{0}
\renewcommand{\thesection}{\Alph{section}}}
\newcommand{\Appendix}[1]{
\refstepcounter{section}
\begin{flushleft}
{\large\bf Appendix \thesection: #1}
\end{flushleft}}
\usepackage{epsfig}

\newcommand\blank[1]{#1}
\renewcommand\blank[1]{}

\newcommand{\Tr}{\operatorname{Tr}}

\def\Bu{{\boldsymbol u}}

\def\Bw{{\boldsymbol w}}

\def\Bv{{\boldsymbol v}}

\def\B0{{\boldsymbol 0}}
\def\BF{{\boldsymbol F}}
\def\BK{{\boldsymbol K}}
\def\BX{{\boldsymbol X}}
\def\BZ{{\boldsymbol Z}}

\def\BOmega{{\boldsymbol\Omega}}
\def\Bvarpi{{\boldsymbol\varpi}}
\def\Bpi{{\boldsymbol\pi}}
\def\CP{{\mathbb C}P}

\def\Tr{{\rm Tr}}

\def\Z{{\bf Z}}

\def\R{{\mathbb R}}
\def\Z{\boldsymbol{Z}}

\newcommand{\BH}{\boldsymbol{H}}

\def\Dbarslash{\,\,{\raise.15ex\hbox{/}\mkern-12mu {\bar D}}}
\def\Dslash{\,\,{\raise.15ex\hbox{/}\mkern-12mu D}}
\def\delslash{\,\,{\raise.15ex\hbox{/}\mkern-9mu \partial}}
\def\delbarslash{\,\,{\raise.15ex\hbox{/}\mkern-9mu {\bar\partial}}}

\def\LAG{\mathscr{L}}

\newcommand{\MAT}[1]{\begin{pmatrix} #1\end{pmatrix}}

\newcommand{\EQ}[1]{\begin{equation}\begin{split} #1
\end{split}\end{equation}}
\newcommand{\AL}[1]{\begin{subequations}\begin{align} #1
\end{align}\end{subequations}}
\newcommand{\SP}[1]{\begin{equation}\begin{split} #1
\end{split}\end{equation}}

\title{The AdS${}_{\boldsymbol{5}}\boldsymbol{\times}$S${}^{\boldsymbol{5}}$ Semi-Symmetric Space Sine-Gordon Theory}
\author{Timothy J. Hollowood\\
Department of Physics,\\ University of Wales Swansea,\\
Swansea, SA2 8PP, UK.\\
E-mail: \email{t.hollowood@swansea.ac.uk}}
\author{and J.~Luis Miramontes\\
Departamento de F\'\i sica de Part\'\i culas and IGFAE,\\
Universidad
de Santiago de Compostela\\ 15782 Santiago de Compostela, Spain\\
E-mail: \email{jluis.miramontes@usc.es}}

\abstract{The generalized symmetric space sine-Gordon theories are a series of $1+1$-integrable field theories that are classically equivalent to superstrings 
on symmetric space spacetimes $F/G$. They are formulated in terms of a semi-symmetric space as a gauged WZW model with fermions and a potential term to deform it away from the conformal fixed point.
We consider in particular the case of $PSU(2,2|4)/Sp(2,2)\times Sp(4)$ which corresponds to $AdS_5\times S^5$. We argue that the infinite tower of conserved charges of these theories includes an exotic ${\cal N}=(8,8)$ supersymmetry that is realized in a mildy non-local way at the Lagrangian level. The supersymmetry is associated to a double central extension of the superalgebra $\mathfrak{psu}(2|2)\oplus\mathfrak{psu}(2|2)$ and includes a non-trivial R symmetry algebra corresponding to global gauge transformations, as well as 2-dimensional spacetime translations. We then explicitly construct soliton solutions and show that they carry an internal moduli superspace $\CP^{2|1}\times\CP^{2|1}$ with both 
bosonic and Grassmann collective coordinates. We show how to semi-classical quantize the solitons by writing an effective quantum mechanical system on the moduli space which takes the form of a 
co-adjoint orbit of $SU(2|2)^{\times2}$. The spectrum consists of a tower of massive
states in the short, or atypical, symmetric representations, just as the giant magnon states of the string world sheet theory, although here the tower is truncated.}

\begin{document}

\section{Introduction}

The Symmetric Space Sine-Gordon (SSSG) theories have received recent attention because they are relativistic integrable theories that are classically equivalent, via the Polhmeyer reduction~\cite{Pohlmeyer:1975nb}\footnote{For a recent review see~\cite{Miramontes:2008wt} and references therein}, to the world-sheet theories of strings on symmetric space spacetimes~\cite{Tseytlin:2003ii,Mikhailov:2005qv,Mikhailov:2005sy}. The SSSG theories
for the cases where the symmetric space $F/G$ is either  $S^n$ or $\CP^n$ have been shown to have soliton solutions which are the images of the string giant magnons under the reduction~\cite{Hofman:2006xt,Dorey:2006dq,Chen:2006gea,Gaiotto:2008cg,Grignani:2008is,Abbott:2009um,Hollowood:2009sc}. Recently, the exact S-matrix for the solitons was conjectured for the complex projective spaces in \cite{Hollowood:2010rv}. Symmetric spaces are characterized by a quotient of groups $F/G$ and the existence of an isometry of $F$ whose stability group is $G$.
However, in order to describe the string world sheet theory for $AdS_5\times S^5$ with all the fermionic degrees-of-freedom,
one needs to generalize the SSSG theories to the case where the symmetric space is replaced by a semi-symmetric space ${\cal F}/G$ which is the quotient of a supergroup with an ordinary group associated to a ${\mathbb Z}_4$ automorphism with $G$ the subgroup fixed by the automorphism~\cite{Serg} (see also \cite{Zarembo:2010sg}).\footnote{In this paper we use the common nomenclature where
$Sp(2n)$ has rank $n$. However, in the companion paper we use the nomenclature that $Sp(n)$ has rank $n$.} For $AdS_5\times S^5$, the relevant semi-symmetric space is
\EQ{
\frac FG=\frac{PSU(2,2|4)}{Sp(2,2)\times Sp(4)}\ .
}
The bosonic part of this space is precisely 
\EQ{
AdS_5\times S^5\thicksim \frac{SO(2,4)}{SO(1,4)}\times \frac{SO(6)}{SO(5)}\thicksim
\frac{SU(2,2)}{Sp(2,2)}\times \frac{SU(4)}{Sp(4)}\ ,
}
and it has been shown that the associated 
semi-symmetric space sine-Gordon (SSSSG) 
theory now involves fermions \cite{Grigoriev:2007bu}. The present work will investigate the two particular examples of these theories, the one above which is, of course central to the gauge-gravity correspondence, and the simpler case,
\EQ{
\frac FG=\frac{PSU(1,1|2)}{U(1)\times U(1)}\ ,
}
whose bosonic part is
\EQ{
AdS_2\times S^2\thicksim\frac{SO(2,1)}{SO(1,1)}\times \frac{SO(3)}{SO(2)}
\thicksim \frac{SU(1,1)}{U(1)}\times\frac{SU(2)}{U(1)}\ .
}
The motivation behind the discovery and investigation of the SSSSG theories, is the question of whether the classical equivalence may extend to quantum equivalence \cite{Grigoriev:2007bu,Mikhailov:2007xr} (see also~\cite{Grigoriev:2008jq,Roiban:2009vh,Hoare:2009rq,Hoare:2009fs,Iwashita:2010tg,Hoare:2010fb}). We keep an open mind about this question, but take the view that answering it will require the quantum solution of these theories and this is the problem that we now address. 

It is a key result of \cite{Grigoriev:2007bu} that the SSSSG theories are classically  integrable and they admit a zero-curvature, or Lax representation, whose algebraic setting involve a very particular  affinization of ${\mathfrak f}$, the Lie algebra of the supergroup $F$, built from the ${\mathbb Z}_4$ automorphism. The fact that there is a proper algebraic setting is key for us because it allows us to generalize various constructions that are used to solve the bosonic theories. To start with, we use it to 
prove that these theories have an infinite tower of conserved charges, some of which can be related directly to conserved currents  local in the Lagrangian fields. The construction here generalizes the same analysis for the ordinary, or bosonic, SSSG theories that appears in the companion papers~\cite{Hollowood:2010dt,Hollowood:2011fm}. In particular, we show that the SSSSG theories have Grassmann conserved charges of Lorentz spin $\pm\frac12$ which are Noether symmetries of the action. These symmetries are candidate supersymmetries, although for the $PSU(2,2|4)$ example the symmetries have a non-local action on some of the fields (which explains why they have not been found before).  We then go further and show that in certain cases, including the $PSU(2,2|4)$ and $PSU(1,1|2)$ examples, those charges generate a closed extended SUSY algebra which includes the energy and momentum and, for the $PSU(2,2,|4)$, a non-abelian $SU(2)^{\times4}$ global symmetry which plays the r\^ole of an R-symmetry. For the $PSU(2,2|4)$ case this extended SUSY algebra is a double central extension of the form
\EQ{
{\mathfrak s}=\big(\mathfrak{psu}(2|2)\oplus\mathfrak{psu}(2|2)\big)\ltimes\big({\mathbb R}\oplus\mathbb R\big)\ ,
}
which is very closely
related to the symmetry algebra of the dyonic giant magnons on the string theory side \cite{Beisert:2005tm,Beisert:2006qh,Arutyunov:2008zt}. However, unlike in the string context, here it is a spacetime symmetry that includes the generators corresponding to two-dimensional spacetime translations as the central extension ${\mathbb R}\oplus\mathbb R\sim\partial_+\oplus\partial_-$. Each $\mathfrak{psu}(2|2)$ factor has 8 Grassmann generators and so there are ${\cal N}=(8,8)$ supersymmetries in total.   

We then go on, following the companion paper~\cite{Hollowood:2010dt}, to
construct soliton solutions in the SSSSG theories using the dressing method. What is interesting is that, in this case, it leads to soliton solutions with both bosonic and Grassmann collective coordinates. If the Grassmann coordinates are turned off then the solitons live entirely in the $S^5$ or $S^2$ factor, and so one can view the Grassmann coordinates as arising from the non-compact $AdS_5$ sector. For $PSU(2,2|4)$ these solitons take the form of non-abelian Q-balls and in the string theory side correspond to dyonic giant magnons~\cite{Hollowood:2011fm}.
We shall calculate the charges of the solitons, including the mass, and argue that the solitons carry an internal moduli space which takes the form of a co-adjoint orbit of the SUSY group $SU(2|2)^{\times2}$. 

In the final section, we proceed, once again following \cite{Hollowood:2010dt,Hollowood:2011fm}, to a semi-classical quantization of the solitons by quantizing their moduli space dynamics. The idea here is to allow the collective coordinates to become time dependent and then substitute this into the action and perform the spatial integral. What remains is an effective quantum mechanical theory on the moduli space. The Grassmann coordinates lead to a fermionic Fock space. For the $PSU(2,2|4)$ case, the bosonic part of the moduli space is a co-adjoint orbit of the Grassmann even subgroup $SU(2)^{\times 4}\subset
SU(2|2)^{\times2}$ and leads to a Hilbert space which are particular representations of $SU(2)^{\times4}$. The bosonic and fermionic states match up in such a way that they form the atypical symmetric representations for each $SU(2|2)$. 

Near the completion of this work, there appeared \cite{Goykhman:2011mq} which also discusses the SUSY of the symmetric space sine-Gordon theories. Our approach has some overlap in that we find that off-shell the SUSY transformations also have a non-local component, however the details are different in that we do not need to modify the original theory.

\section{The Semi-Symmetric Space Sine-Gordon Theories}
\label{sssg}

For the case of ordinary groups, the SSSG theories are related to a triplet of groups 
$H\subset G\subset F$, where $F/G$ is the symmetric space in question. 
The group in the numerator $F$ admits 
an involution $\sigma_-$ whose stabilizer is the
subgroup $G$. Acting on the Lie algebra of $F$, the involution gives rise to the canonical decomposition
\EQ{
{\mathfrak f} = {\mathfrak g} \oplus {\mathfrak p}
\quad \text{with} \quad 
[{\mathfrak g},{\mathfrak g}]\subset {\mathfrak g}\>, 
\quad [{\mathfrak g},{\mathfrak p}]\subset 
{\mathfrak p}\>, 
\quad [{\mathfrak p},{\mathfrak p}]\subset {\mathfrak g}\>,
\label{CanonicalDec}
}
where ${\mathfrak g}$ and ${\mathfrak p}$ are the $+1$ and $-1$ eigenspaces of $\sigma_-$, respectively.
This structure gives rise to the loop, or affine, algebra
\EQ{
\hat{\mathfrak f}= \bigoplus_{n\in\Z} \left(z^{2n} \otimes {\mathfrak g}
  +z^{2n+1} \otimes {\mathfrak p} \right)\ ,
 }
which plays an important role in the study of these theories.
The SSSG equations can be written in Lax form as a zero curvature
condition for a connection that depends on an auxiliary complex parameter $z$, which is a spectral parameter. Said another way, the connection is actually valued in the affine Lie algebra $\hat{\mathfrak f}$. More explicitly, the SSSG equations are
\EQ{
[{\cal L}_\mu(z),{\cal L}_\nu(z)]=0\ ,
\label{zcc}
}
where
\SP{
{\cal L}_+(z)&= \partial_++\gamma^{-1}\partial_+\gamma+\gamma^{-1}A_+\gamma-z \Lambda
\ ,\\[5pt]
{\cal L}_-(z)&= \partial_-+A_--z^{-1}\gamma^{-1}\Lambda\gamma\ .
\label{lxc}
}
Here,  $\Lambda$ is an element of the the $-1$
eigenspace $\mathfrak p$ of the Lie algebra of $F$ under $\sigma_-$, and $\gamma$ takes values in $G$.
The gauge connection $A_\pm$ is associated to the vector-like gauge symmetry
generated by the subgroup $H\subset G$ which commutes with $\Lambda$:
\EQ{
\gamma\longrightarrow U\gamma U^{-1}\ ,\qquad A_\pm\longrightarrow U\big(A_\pm+\partial_\pm\big)U^{-1}\ , \qquad U\in H\ .
\label{gaugeLR2}
}
The Lagrangian formulation of the SSSG equations was originally proposed in~\cite{Bakas:1995bm}. They arise as the Lagrange equations of the action
\SP{
S[\gamma,A_\mu]=S_\text{gWZW}[\gamma,A_\mu]-\frac k{\pi}\int d^2x\,\Tr\Big(
\Lambda
\gamma^{-1}\Lambda\gamma\Big)\ ,
\label{ala}
}
where $S_\text{gWZW}[\gamma,A_\mu]$ is the usual WZW action 
for the gauged WZW model for $G/H$, and the integer number $k$ is the level.

The structure above has been generalized to the case of a semi-symmetric space
$F/G$ in \cite{Grigoriev:2007bu,Grigoriev:2008jq}, and here we shall consider the examples
\EQ{
\frac{PSU(2,2|4)}{Sp(2|2)\times Sp(4)}\qquad\text{and}\qquad\frac{PSU(1,1|2)}{U(1)\times U(1)}\,.
}
The numerator group $F$ is now a supergroup, and the r\^ole of the involution is played by a ${\mathbb Z}_4$ automorphism $\sigma_-$. Our conventions for superalgebras are  
taken from \cite{Arutyunov:2009ga}. First of all, the superalgebra ${\mathfrak sl}(2N|2N)$ is defined by the $4N\times4N$ matrices
\EQ{
M=\MAT{m&\theta\\ \eta&n}\ ,
}
where $m$ and $n$ are Grassmann even and $\theta$ and $\eta$ are Grassmann odd. 
These matrices are required to have vanishing supertrace\footnote{Notice that our convention is the opposite of \cite{Grigoriev:2007bu}, so that the supertrace is positive on the $S^5$ factor and negative on the $AdS_5$ factor.}
\EQ{
\text{STr}\,M=-\Tr\,m+\Tr\,n=0\ .
\label{STr}
}
The non-compact real form ${\mathfrak su}(N,N|2N)$ is picked out by imposing the reality condition
\EQ{
M=-HM^\dagger H
\label{real}
}
where, in $N\times N$ block form,
\EQ{
H=\left(\begin{array}{cc|cc} {\mathbb I}_N &&&\\ &-{\mathbb I}_N &&\\ \hline &&{\mathbb I}_N&
\\ &&&{\mathbb I}_N\end{array}\right)\,.
}
Here, $\dagger$ is the usual hermitian conjugation, $M^\dagger=(M^*)^t$, but with the definition that complex conjugation is anti-linear on products of Grassmann odd elements
\EQ{
(\theta_1\theta_2)^*=\theta_2^*\theta_1^*\ ,
}
which guarantees that $(M_1M_2)^\dagger=M_2^\dagger M_1^\dagger$.
The superalgebra ${\mathfrak{psu}}(N,N|2N)$ is then the quotient of $\mathfrak{su}(N,N|2N)$ by the unit element $i{\mathbb I}_{4N}$, which is a centre of the algebra. 

For the cases of interest $N=1,2$, the ${\mathbb Z}_4$ autormorphism is defined as
\EQ{
M\longrightarrow\sigma_-(M)=-{\cal K}M^{st}{\cal K}^{-1}\ ,
}
where $st$ denotes the ``super-transpose'' defined as
\EQ{
M^{st}=\MAT{m^t&-\eta^t\\ \theta^t&n^t}\ .
}
For the case $\mathfrak{psu}(2,2|4)$
\EQ{
{\cal K}=\left(\begin{array}{cc|cc} J_2 &  &  & \\
  & J_2 &  & \\ \hline
 &  & J_2 & \\
 &  & & J_2\end{array}\right)\ ,\qquad J_2=\MAT{0&-1\\ 1&0}\ ,
}
while for $\mathfrak{psu}(1,1|2)$
\EQ{
{\cal K}={\mathbb I}_4\ .
 }
 Under $\sigma_-$, the superalgebra ${\mathfrak{psu}}(N,N|2N)$ has the decomposition
\EQ{
{\mathfrak f}= {\mathfrak f}_0\oplus{\mathfrak f}_1\oplus{\mathfrak f}_2\oplus{\mathfrak f}_3\ ,\qquad \sigma_-({\mathfrak f}_j)=i^j\,{\mathfrak f}_j\ , \qquad [{\mathfrak f}_j,{\mathfrak f}_k]\subset {\mathfrak f}_{j+k\; \text{mod}\; 4}\,.
\label{CanonicalDecN}
}
In particular, the even graded parts are Grassmann even while the odd graded parts are Grassmann odd. The zero graded part ${\mathfrak f}_0\equiv{\mathfrak g}$ is the (bosonic) Lie algebra
of $G$, which is the group in the denominator of the semi-symmetric space. For our two examples, $G$
equals $Sp(2,2)\times Sp(4)$ and $U(1)\times U(1)$, for $N=2$ and $N=1$, respectively. Correspondingly, ${\mathfrak f}_0\oplus{\mathfrak f}_2$ is the Lie algebra of $SU(N,N)\times SU(2N)$, which is the bosonic subgroup of $F$.
The fermionic parity is defined by
\EQ{
M\longrightarrow{\mathfrak P}M{\mathfrak P}=\MAT{m&-\theta\\ -\eta&n}\ ,\qquad
{\mathfrak P}=\left(\begin{array}{c|c} -{\mathbb I}_{2N} & 0 \\ \hline
 0 & {\mathbb I}_{2N}\end{array}\right)\ .
}

Like its bosonic cousin, the generalized
SSSSG theory is associated to a loop algebra, which in this case is the graded affine algebra
\EQ{
\hat{\mathfrak f}= \bigoplus_{n\in\Z} \bigoplus_{j=0}^3 z^{4n+j} \otimes {\mathfrak f}_j\equiv \bigoplus_{k\in\Z}\hat{\mathfrak f}_k\,,
 \label{NewAffine}
 }
where we have defined $\hat{\mathfrak f}_k = z^k\otimes {\mathfrak f}_j$ for $k-j\in 4\Z$, and $[\hat{\mathfrak f}_k,\hat{\mathfrak f}_l]\subset \hat{\mathfrak f}_{k+l}$ as a consequence of the decomposition~\eqref{CanonicalDecN}.
It will become apparent as we proceed that the grade of an element is twice its Lorentz spin, where $x^\pm$ are assigned spin $\mp1$.
The Lax connection, generalizing \eqref{lxc}, takes the form\footnote{Our conventions are a trivial re-labelling of those of \cite{Grigoriev:2007bu}, $\psi_+=\Psi_R$, $\psi_-=\Psi_L$ and $\Lambda=-T$. We also take the mass parameter $\mu=1$ because it can always be re-instated by dimensional analysis.}
\SP{
{\cal L}_+(z)&= \partial_++\gamma^{-1}\partial_+\gamma+\gamma^{-1}A_+\gamma+
z\psi_+-z^2\Lambda
\ ,\\[5pt]
{\cal L}_-(z)&= \partial_-+A_-+z^{-1}\gamma^{-1}\psi_-\gamma-z^{-2}\gamma^{-1}\Lambda\gamma\ ,
\label{susylax}
}
where $\gamma\in G$, and $\psi_\pm$ are fields taking values in ${\mathfrak f}_{1,3}$, respectively, and hence are fermionic. The SSSSG equations are relativistically invariant, and the form of the Lax connection exhibits that the Lorentz transformation
$x^\pm\to \lambda^{\mp1} x^\pm$ is equivalent to the rescaling of the spectral parameter $z\to \lambda^2 z$.

The theory is defined by the choice of the constant element $\Lambda\in\mathfrak f_2$. This in turn is determined from the string world sheet theory to be~\cite{Grigoriev:2007bu}
\footnote{
In fact, we can think of the possible choices for $\Lambda$ in the following alternative way.  If we turn off all the fermionic fields, then the bosonic fields are associated to the product of two ordinary symmetric spaces, that is $SU(2,2)/Sp(2,2)$ and $SU(4)/Sp(4)$ for the $PSU(2,2|4)$ example. Then the possible choices for $\Lambda$ in each of the corresponding bosonic theories are determined by the rank of the symmetric spaces and by their signature~\cite{Miramontes:2008wt}. 
In the present case, $SU(4)/Sp(4)=S^5$ has rank 1 and definite signature, and so there is a unique choice  up to conjugation, say $\Lambda_2$, with $\Tr(\Lambda_2^2)<0$. Then we have
$\Lambda=\mu_1\Lambda_1+\mu_2\Lambda_2$, and the Virasoro constraints require
\EQ{
\text{STr}\,\Lambda^2 = -\mu_1^2 \Tr(\Lambda_1^2) + \mu_2^2 \Tr(\Lambda_2^2)=0\,.
}
Since $SU(2,2)/Sp(2,2)=AdS_5$ has also of rank 1 but is of indefinite signature,  there are two possible solutions. The first one is $\Tr(\Lambda_1^2)=0$ with $\mu_1\not=0$  and $\mu_2=0$. The second is  $\Tr(\Lambda_1^2)<0$ with $\mu_1, \mu_2\not=0$, which is the one considered in~\cite{Grigoriev:2007bu}, and the one that corresponds to~\eqref{Lambda}.}
\EQ{
\Lambda=\frac i2\left(\begin{array}{cc|cc} {\mathbb I}_N &&&\\ &-{\mathbb I}_N &&\\ \hline &&{\mathbb I}_N&
\\ &&&-{\mathbb I}_N\end{array}\right)\ .
\label{Lambda}
}
An important point for what follows is that the choice for $\Lambda$ is {\it semi-simple\/}, meaning that the algebra ${\mathfrak f}$
has a decomposition
\EQ{
{\mathfrak  f}=\text{Ker}(\text{ad}\,\Lambda)\oplus
  \text{Im}(\text{ad}\,\Lambda)\equiv {\mathfrak
  f}^\perp\oplus{\mathfrak f}^\parallel\ ,
\label{dec}
}
which lifts to the affine algebra $\hat{\mathfrak f}$.
In the present cases, there is the additional simplifying feature that
\SP{
[{\mathfrak f}^\perp,{\mathfrak f}^\perp]={\mathfrak f}^\perp\ ,\qquad
[{\mathfrak f}^\perp,{\mathfrak f}^\parallel]={\mathfrak f}^\parallel\ ,\qquad
[{\mathfrak f}^\parallel,{\mathfrak f}^\parallel]={\mathfrak f}^\perp\ ;
\label{Orthogonal2}
}
and so the decomposition \eqref{dec} gives an alternative  ${\mathbb Z}_2$ gradation of ${\mathfrak  f}$ which we denote
\EQ{
\tau(\mathfrak f^\perp)=\mathfrak f^\perp\ ,\qquad  
\tau(\mathfrak f^\parallel)=-\mathfrak f^\parallel\ .
}
In particular, it is useful to note that
\EQ{
\Lambda{\mathfrak f}^\perp={\mathfrak f}^\perp\Lambda\ ,\qquad
\Lambda{\mathfrak f}^\parallel=-{\mathfrak f}^\parallel\Lambda\ .
}
The projectors onto the subspaces can be written as
\EQ{
{\cal P}^\perp=-\{\Lambda,\{\Lambda,\cdot\}\}\ ,\qquad
{\cal P}^\parallel=-[\Lambda,[\Lambda,\cdot]]\ .
}

In a way that will be uncovered later, the affinization of the subalgebra $\hat{\mathfrak f}^\perp$ plays an important r\^ole as a symmetry algebra. In patricular the zero graded component $\mathfrak h\equiv\hat{\mathfrak f}^\perp_0={\mathfrak f}^\perp_0$ generates an ordinary Lie group $H\subset G$. For the $PSU(1,1|2)$ example $H=\emptyset$, while for the  $PSU(2,2|4)$ example $H=SU(2)^{\times4}$ which is schematically of the form
\EQ{
\left(\begin{array}{cc|cc} SU(2)_f^{(+)} &&&\\ &SU(2)_f^{(-)} &&\\ \hline &&SU(2)_b^{(+)}&
\\ &&&SU(2)_b^{(-)}\end{array}\right)\ ,
}
where $SU(2)\simeq Sp(2)$. The labels on the $SU(2)$ subgroups uniquely identify them and will be needed later. The group $H$ is precisely the subgroup of $G$  that is gauged in the Lagrangian formulation described below and, in~\eqref{susylax}, $A_\pm \in{\mathfrak h}$. For the $PSU(1,1|2)$ example, notice that no gauging will be required.

In terms of the component fields, the equations-of-motion $[{\cal L}_+(z), {\cal L}_-(z)]=0$ are
\AL{
&\partial_-(\gamma^{-1}\partial_+\gamma+\gamma^{-1}A_+\gamma)-\partial_+A_-+[A_-,\gamma^{-1}\partial_+\gamma+\gamma^{-1}A_+\gamma]\notag\\  & \qquad\qquad
-[\psi_+,\gamma^{-1}\psi_-\gamma]-[\Lambda,\gamma^{-1}\Lambda\gamma]=0\ ,\label{eqo}\\[5pt]
&D_\mp\psi_\pm+ [\Lambda,\gamma^{\mp1}\psi_\mp\gamma^{\pm1}]=0\ .
\label{fem}
}
These equations-of-motion follow from an action of the form
\SP{
S&=S_\text{gWZW}[\gamma,A_\mu]-\frac k{\pi}\int d^2x\,\text{STr}\,\Big(\Lambda
\gamma^{-1}\Lambda\gamma\Big)\\
&+\frac k{2\pi}\int d^2x\,\text{STr}\,\Big(\psi_+[\Lambda,D_-\psi_+]-\psi_-[\Lambda,D_+\psi_-]
-2\psi_+\gamma^{-1}\psi_-\gamma\Big)\ ,
\label{alp}
}
where $S_\text{gWZW}[\gamma,A_\mu]$ is the conventional (bosonic) gauged WZW model for $G/H$ with level $k$, but involving the supertrace~\eqref{STr} rather than the ordinary trace. 
It is invariant under the gauge transformations
\EQ{
\gamma\longrightarrow U\gamma U^{-1}\,,\qquad
A_\mu\longrightarrow U\big(A_\mu+\partial_\mu\big)U^{-1}\,,\qquad U\in H\ .
}
An important feature of these theories is that they admit soliton solutions whose fields do not fall-off at $x=\pm\infty$, and it was emphasized in~\cite{Hollowood:2010dt,Hollowood:2011fm} that this makes the WZ term require careful treatment. In particular, it cannot strictly speaking be defined as an integral over a three-dimensional space with the two-dimensional spacetime as a boundary. One way to unambiguously define the action is, as in~\cite{Hoare:2009fs}, to use the condition of gauge invariance to pin down the expansion of the WZ term in terms of $\phi$, with $\gamma=e^\phi$. This prescription requires to supplement the action with a boundary term
\EQ{
-\frac{k}{2\pi} \int d^2x\> \epsilon^{\mu\nu}\partial_\mu \Tr\big(A_\nu \phi \big)\ ,
\label{Topol}
}
which does not contribute to the equations of motion.
The gauge transformations of the fermionic variables
read as follows
\SP{
\psi_\pm \longrightarrow U \psi_\pm U^{-1}\,.
}
Moreover, in the action above it is assumed that the fermionic fields satisfy the constraints
\EQ{
\psi_\pm^\perp=0\ ,
\label{kapf}
}
Notice that these conditions are consistent with the equations-of-motion 
\eqref{fem}, and it turns out that they arise very naturally in the string theory setting
as fixing the residual $\kappa$-symmetry of the world-sheet theory \cite{Grigoriev:2007bu}.

The equations-of-motion of the gauge field imply the vanishing of the (na\"\i ve) gauge current on-shell\footnote{Where necessary, the notation $\approx$ will indicate equality on-shell.}
\EQ{
J_\pm=\pm\Big(\gamma^{\mp1}\partial_\pm\gamma^{\pm1}+\gamma^{\mp1}A_\pm\gamma^{\pm1}\Big)^\perp\mp A_\pm\pm2\Lambda\psi_\pm\psi_\pm\approx0\ .
\label{gcond}
}
The equations \eqref{eqo} projected onto $\hat{\mathfrak f}^\perp$, together with \eqref{gcond}, imply the flatness condition
\EQ{
[\partial_++A_+,\partial_-+A_-]\approx 0\ ,
\label{fns}
}
and so one can fix the gauge---at least on-shell---by taking $A_\mu=0$, and from \eqref{gcond} we have the constraints
\EQ{
\big(\gamma^{\mp1}\partial_\pm\gamma^{\pm1}\big)^\perp+2\Lambda\psi_\pm\psi_\pm=0\ .
\label{cons}
}
A convenient gauge to choose off-shell is a kind of light-cone gauge which imposes the Lorentz invariance condition $A_+=0$ \cite{Hoare:2009rq,Hoare:2009fs,Hoare:2010fb}. Anyway, for the solutions to the equations-of-motion, we will always be able to take $A_\mu=0$. Note that all the complications of the gauge fields are absent in the simpler $PSU(1,1|2)$ case where $H$ is trivial.

Although the gauge current in \eqref{gcond} vanishes, we expect that physical configurations will carry charge under the global subgroup of the gauge group. This seems paradoxical, but actually it is typical of a gauge theory and it is well known that the charge
charge does not come from integrating the na\"\i ve temporal component of the gauge current. In fact, as we have seen above, this current vanishes on-shell. Rather the physical current receives a ``topological" contribution that is fixed by the boundary term~\eqref{Topol}. Writing $\gamma=e^\phi$, the true Noether current is~\cite{Hollowood:2010dt,Hollowood:2011fm}
\EQ{
{\cal J}^\mu=J^\mu+\epsilon^{\mu\nu}\partial_\nu\phi^\perp\ ,
\label{kss}
}
which is sensitive to the behaviour of the field at spatial infinity. On-shell, the charge is then equal to a kink charge
\EQ{
{\cal Q}=\int dx\, {\cal J}^0\approx-\int dx\,\partial_1\phi^\perp=
-\phi^\perp(\infty)+\phi^\perp(-\infty)=q_0\ .
\label{chgs1}
}
Note that at $x=\pm\infty$ the group field must lie in a minimum of the potential so that $\phi(\pm\infty)\in\mathfrak h$ and consequently the projection onto $\mathfrak h=\mathfrak g^\perp$ is unnecessary. 
Assuming that $\gamma(\pm\infty)$ commute, which will be true for the configurations that we consider, it follows that the kink charge corresponds precisely to
\EQ{
\gamma(\infty)^{-1}\gamma(-\infty)=e^{q_0}\ .
\label{chgs2}
}

\section{The Integrable System}\label{iss}

Underlying the SSSSG theories there is a new kind of integrable system which is based on the graded affine algebra $\hat{\mathfrak f}$. The system leads to a set of what are generally understood as hidden symmetries of the SSSSG theory generated by the elements of the subalgebra $\hat{\mathfrak f}^\perp$. An important part of our story is that these symmetries include a finite subalgebra which has the form of an extended supersymmetry algebra, and which they are
conventional Noether symmetries associated to local conserved currents.

The Lax equations \eqref{zcc}
are the integrability conditions for the associated linear problem
\EQ{
{\cal L}_\mu(z) \Upsilon(z)=0\ .
\label{LinProb}
}
It follows that the ``subtracted monodromy"~\cite{Hollowood:2010dt}
\EQ{
{\cal M}(z)=\lim_{x\to\infty}\Upsilon_0(x;z)^{-1}\Upsilon(x;z)\Upsilon^{-1}(-x;z)\Upsilon_0(-x;z)\ ,
\label{MMatrix2}
}
is constant in time, where the subtraction involves removing the effects of 
\EQ{
\Upsilon_0(x;z)=\exp\big[(z^2x^++z^{-2}x^-)\Lambda\big]\ ,
\label{dvs}
}
which is the vacuum solution of the linear problem ($\gamma=1$, $\psi_\pm=0$, $A_\mu=0$).
Note that individually the quantities $\Upsilon(x;z)$ and $\Upsilon_0(x;z)$ diverge as $x\to\pm\infty$, but the subtracted monodromy is a finite quantity.

The expansion of the subtracted monodromy around $z=0$ and $\infty$ provide an set of conserved charges $q_s$,
\EQ{
{\cal M}(z)=\exp\big[
q_0+q_{1}z+q_{2}z^2+\cdots\big]=\exp\big[q_{-1}/z+q_{-2}/z^2+\cdots\big]\ ,
\label{mex}
}
of Lorentz spin $\tfrac s2$,  and we will soon show that $q_s\in{\mathfrak f}^\perp$. Generally, these charges are non-local quantities, however, some of them are associated to local conserved currents. This includes the spin $\pm\frac12$ charges $q_{\pm1}$ as well as the components of $q_s$ along the centre of ${\mathfrak f}^\perp$, which includes the infinite set of elements $z^{2+4n}\Lambda$ with $n\in\Z$.
In particular, the conserved 2-momentum of a configuration is given by
\EQ{
p_\pm=\mp\frac {k}{2\pi}\text{STr}\big(\Lambda q_{\mp2}\big)\ .
\label{tmn}
}

The form of the conserved currents can be deduced using the Drinfeld-Sokolov procedure~\cite{DeGroot:1991ca}. In the following, we will impose the on-shell gauge $A_\mu=0$ and start by considering the currents of positive spin. To this end, we introduce
\EQ{
\Phi(z)=\exp\,y(z)\>,\qquad y(z)=\sum_{s\geq1}y_{-s}\,z^{-s} \in \hat{\mathfrak f}_{<0}
}
and solve (off-shell, up to the choice of gauge fixing conditions)
\EQ{
\Phi(z)^{-1} {\cal L}_+(z) \Phi(z)=  \partial_+  - z^2\Lambda+ h_+(z)\,,\qquad
h_+(z)=\sum_{s\leq 1}h_{s,+}\,z^{s}\in\hat{\mathfrak f}^\perp_{\leq 1}\ ,
\label{DS}}
which can be done order-by-order in $z$ as we illustrate below.
Correspondingly,
\EQ{
\Phi(z)^{-1} {\cal L}_-(z)\Phi(z) = \partial_-  +h_-(z)\,, \qquad h_-(z)\in \hat{\mathfrak f}_{\leq-1}\,.
\label{DS2}
}
Then, the zero curvature condition~\eqref{zcc} becomes (on-shell)
\EQ{
\bigl[\partial_+ -z^2\Lambda+ h_+(z),  \partial_-  + h_-(z)\bigr]=0\;\; \Rightarrow \;\; h_-(z)\in \hat{\mathfrak f}^\perp_{\leq-1}\,.
\label{Zero2}
}
which implies, in particular, that
\EQ{
\partial_- h_{1,+}=0\,,\qquad
\partial_- h_{0,+}=[h_{1,+}, h_{-1,-}]\,.
}
This shows that we can consistently set
\EQ{
h_{1,+}=h_{0,+}=0\;\; \Rightarrow\;\; h_+(z)\in \hat{\mathfrak f}^\perp_{\leq -1}\,.
\label{newConst}
}
This type of constraints are well known in the context of integrable hierarchies~\cite{Hollowood:1992ku,Miramontes:1998hu,Madsen:1999ta}.
Below, we will show that they are equivalent to~\eqref{kapf} and~\eqref{cons}.
In addition, the zero-curvature condition~\eqref{Zero2}
implies that the components of $h_\pm(z)$ in the centre of $\hat{\mathfrak f}^\perp$ lead directly to conserved currents. 

In~\cite{Miramontes:1998hu}, it was emphasized that the choice of $\Phi$ is not unique. It is defined modulo the transformations $\Phi\rightarrow \Phi\eta$ with
\EQ{
\eta \in \exp \,\hat{\mathfrak f}^\perp_{<0}\>,
}
which do not change the form of~\eqref{DS}---but do change the value of $h_\pm(z)$. However,  the solution can always be chosen such that $\Phi$ and $h_\pm(z)$ are local functions of the component fields by simply enforcing the condition
\EQ{
y(z)\in  \hat{\mathfrak f}^\parallel_{<0}\,,
\label{LocalCond}
}
which will be used in the following. This can be proved by induction as will become clear as we show the beginning of this process below.
At order $z$, $z^0$ and  $z^{-1}$, \eqref{DS} gives
\AL{
h_{1,+} -[y_{-1},\Lambda]&=\psi_+\ ,\label{Recurr1}\\[5pt]
h_{0,+}- [y_{-2},\Lambda]&=\gamma^{-1}\partial_+\gamma-[y_{-1},\psi_+]-\tfrac12[y_{-1},[y_{-1},\Lambda]]\ ,\label{Recurr2}\\[5pt]
h_{-1,+}-[y_{-3},\Lambda] &=\partial_+ y_{-1} -[y_{-1},\gamma^{-1}\partial_+\gamma]-[y_{-2},\psi_+] \notag\\
&~~~~+\tfrac12[y_{-1},[y_{-1},\psi_+]]-\tfrac{1}{2}[y_{-1},[y_{-2},\Lambda]]\notag\\ &~~~~ -\tfrac{1}{2}[y_{-2},[y_{-1},\Lambda]] +\tfrac16[y_{-1},[y_{-1},[y_{-1},\Lambda]]]\ ,
\label{Recurr3}
}
and \eqref{DS2} at order $z^{-1}$ provides
\AL{
h_{-1,-}&=\partial_-y_{-1}+\gamma^{-1}\psi_-\gamma\ .
\label{Recurr5}
}

The first equation \eqref{Recurr1} shows that
\EQ{
h_{1,+}= \psi_+^\perp\,,
}
which identifies the first constraint in~\eqref{newConst} with the $\kappa$-symmetry fixing condition \eqref{kapf}.
Then, using the expressions for the projectors onto $\hat{\mathfrak f}^\perp$ and $\hat{\mathfrak f}^\parallel$, \eqref{Recurr1} is solved by
\EQ{
[y_{-1},\Lambda]=-\psi_+ \qquad
\implies\qquad y_{-1}=[\psi_+,\Lambda]\ .
}
Projecting \eqref{Recurr2} onto ${\mathfrak f}^\perp$, 
and using \eqref{cons}, we have
\EQ{
h_{0,+}= (\gamma^{-1}\partial_+\gamma)^\perp+2\Lambda\psi_+\psi_+=0\>,
\label{hzp}
}
which is the second constraint in~\eqref{newConst}.
Hence, there is no zero-graded conserved current which is a reflection of the vanishing of the of the na\"\i ve gauge current \eqref{gcond} on-shell.
Projecting  \eqref{Recurr2} onto ${\mathfrak f}^\parallel$ determines $y_{-2}$:
\EQ{
[y_{-2},\Lambda]=-(\gamma^{-1}\partial_+\gamma)^\parallel\qquad
\implies\qquad y_{-2}=[(\gamma^{-1}\partial_+\gamma)^\parallel,\Lambda]\ .
}
Moving on to the next level \eqref{Recurr3} and projecting onto $\hat{\mathfrak f}^\perp$ gives
\EQ{
h_{-1,+}&=-[y_{-1},(\gamma^{-1}\partial_+\gamma)^\parallel]-[y_{-2},\psi_+]\\[5pt]
&~~~~-\tfrac12[y_{-1},[y_{-2},\Lambda]]
-\tfrac12[y_{-2},[y_{-1},\Lambda]]\\[5pt]
&=[[\Lambda,(\gamma^{-1}\partial_+\gamma)^\parallel],\psi_+]
\ .
}
In addition, from  \eqref{Recurr5} we have
\EQ{
h_{-1,-}=\big(\gamma^{-1}\psi_-\gamma)^\perp\ .
}

From the zero curvature condition \eqref{Zero2}, and using the constraints \eqref{newConst}, 
the quantities $h_{-1,\pm}$ imply the existence of a 
spin $\tfrac32$ conserved current ${\cal G}^{(+)}_\mu$
\EQ{
{\cal G}^{(+)}_+=-h_{-1,+}\ ,\qquad {\cal G}^{(+)}_-=h_{-1,-}\ ,\qquad \partial^\mu{\cal G}^{(+)}_\mu=0\ .
\label{gcur}
}
The associated conserved charge is
\AL{
q_{-1}&=\frac12\int_{-\infty}^\infty dx\,\Big\{
\tfrac12[[\Lambda,\psi_+],(\gamma^{-1}\partial_+\gamma)^\parallel]+\tfrac12[[\Lambda,\gamma^{-1}\partial_+\gamma],\psi_+]-(\gamma^{-1}\psi_-\gamma)^\perp\Big\}\\[5pt]
&=\frac12\int_{-\infty}^\infty dx\,\Big\{
[[\Lambda,\gamma^{-1}\partial_+\gamma],\psi_+]-(\gamma^{-1}\psi_-\gamma)^\perp\Big\}
\ ,
\label{fms1}
}
of spin $\tfrac12$. 

In a similar way, a second set of conserved densities with negative spin can be constructed starting from 
\EQ{
\tilde{\cal L}_-(z)\equiv \gamma{\cal L}_-(z)\gamma^{-1}&= \partial_- -\partial_-\gamma\gamma^{-1}+ +z^{-1}\psi_- - z^{-2}\Lambda\ ,\\
\tilde{\cal L}_+(z)\equiv\gamma{\cal L}_+(z)\gamma^{-1}&=\partial_++z\gamma\psi_+\gamma^{-1}-z^2\gamma\Lambda\gamma^{-1}\ ,
\label{DSMinus}
}
instead of ${\cal L}_\pm$, with 
\EQ{
\Phi\rightarrow \tilde\Phi\in \exp \hat{\mathfrak f}^\parallel_{>0}\ ,\quad
h_\mu(z)\rightarrow \tilde h_\mu(z)=\sum_{s> 0} h_{s,\mu}z^s\in \hat{\mathfrak f}^\perp_{>0}\ .
}
The two quantities $h(z)$ and $\tilde h(z)$ are trivially related by means of the parity transformation
\EQ{
z\Lambda\rightarrow z^{-1}\Lambda,\quad
\partial_+\rightarrow \partial_-,\quad
\gamma\rightarrow \gamma^{-1},\quad
\psi_\pm\to\psi_\mp\ .
\label{Parity}
}
In particular, this provides a conserved current ${\cal G}^{(-)}_\mu$ with charge
$q_{1}$  of spin $-\tfrac12$:
\AL{
q_{1}&=\frac12\int_{-\infty}^\infty dx\,\Big\{
\tfrac12[[\Lambda,\psi_-],(\gamma\partial_-\gamma^{-1})^\parallel]+\tfrac12[[\Lambda,\gamma\partial_-\gamma^{-1}],\psi_-]-(\gamma\psi_+\gamma^{-1})^\perp\Big\}\\[5pt]
&=\frac12\int_{-\infty}^\infty dx\,\Big\{
[[\Lambda,\gamma\partial_-\gamma^{-1}],\psi_-]-(\gamma\psi_+\gamma^{-1})^\perp\Big\}\ .
\label{fms2}
}
Later in this section, we will address the question of whether the spin $\pm\frac12$ charges $q_{\mp1}$ with associated currents ${\cal G}^{(\pm)}_\mu$ are actually evidence of an underlying SUSY. 

The procedure for calculating the densities quickly becomes very involved and we will not pursue it any further. However, at the next level $\text{STr}(\Lambda h_{\pm 2,\mu})$ give the components of the energy-momentum tensor, much as in the bosonic theories \cite{Hollowood:2010dt},
\EQ{
&T_{++}=\frac k{2\pi}\text{STr}\,(\Lambda h_{-2,+})\ ,\qquad~~ T_{-+}=-\frac k{2\pi}\text{STr}\,(\Lambda h_{2,-})\ ,\\
&T_{--}=\frac k{2\pi}\text{STr}\,(\Lambda h_{2,-})\ ,\qquad T_{+-}=-\frac k{2\pi}\text{STr}\,(\Lambda h_{-2,+})\ .
}
Moreover, it is clear by induction that by imposing the condition \eqref{LocalCond}, $y_s$ is determined locally in terms of the fields and their derivatives as claimed above.

In order to deduce the relationship between the conserved densities and the subtracted monodromy~\eqref{MMatrix2}, we solve the zero curvature condition~\eqref{Zero2} as follows
\EQ{
h_+(z)= \Omega\partial_+ \Omega^{-1}\>,\qquad
h_-(z)= -z^{-2}\Lambda+ \Omega\partial_- \Omega^{-1}\>,
\qquad
\Omega\in \exp \hat{\mathfrak f}^\perp_{<0}\ .
}
This leads to
\EQ{
\chi^{-1}{\cal L}_\pm(z)\chi=\partial_\pm -z^{\pm2} \Lambda\>,\qquad
\chi = \Phi\Omega \in \exp \hat{\mathfrak f}_{<0}\ .
\label{MiniLax1}
}
In other words, $\chi\equiv\chi(z)$ is a formal series in $z^{-1}$ taking values in $F$ normalized such that $\chi=1$ at $z=\infty$. This provides the following expression for the solution to the associated linear problem~\eqref{LinProb}: 
\EQ{
\Upsilon(z)= \chi(z) \Upsilon_0(z) g_+,
\label{LPsol1}
}
where $\Upsilon_0(z)$ is the vacuum solution defined in \eqref{dvs}, and
$g_+$ is a constant element of the loop group associated to $\hat{\mathfrak f}$.
In a completely analogous fashion, starting from $\tilde{\cal L}_\pm(z)$ instead of ${\cal L}_\pm(z)$ we get
\EQ{
\tilde\chi^{-1}\tilde{\cal L}_\pm(z)\tilde\chi=\partial_\pm -z^{\pm2} \Lambda\>,\qquad
\tilde\chi=  \tilde\Phi\tilde\Omega\in \exp \hat{\mathfrak f}_{>0}\>,
\label{MiniLax2}
}
where
\EQ{
\tilde h_+(z)= -z^2\Lambda +\tilde \Omega\partial_+ \tilde\Omega^{-1}\>,\qquad
\tilde h_-(z)= \tilde\Omega\partial_-\tilde\Omega^{-1}\>.
}
In this case, $\tilde\chi\equiv\tilde\chi(z)$ is a formal series in $z$ normalized such that $\tilde\chi(0)=1$, and~\eqref{MiniLax2} provides a different expression for the solution to the associated linear problem:
\EQ{
\Upsilon(z)= \gamma^{-1} \tilde\chi(z) \Upsilon_0(z) g_-,
\label{LPsol2}
}
where $g_-$ is another constant element of the loop group associated to $\hat{\mathfrak f}$. Equating~\eqref{LPsol1} and~\eqref{LPsol2} gives rise to the factorization (Riemann-Hilbert) problem
\EQ{
\Upsilon_0(z) g_-g_+^{-1} \Upsilon_0^{-1}(z)=\tilde\chi(z)^{-1}\gamma\chi(z)\>.
\label{RHp}
}

Equations,~\eqref{LPsol1} and~\eqref{LPsol2} lead to two different expression for the subtracted monodromy:
\EQ{
&{\cal M}(z)\\
&=\lim_{ x\to\infty}\Upsilon^{-1}_0( x;z)\chi( x;z)
\Upsilon_0( x;z)\Upsilon^{-1}_0(- x;z)\chi^{-1}(- x;z)
\Upsilon_0(- x;z)\ ,\\[5pt]
&=\lim_{ x\to\infty}\Upsilon^{-1}_0( x;z)\gamma^{-1}( x)\tilde\chi( x;z)
\Upsilon_0( x;z)\Upsilon^{-1}_0(- x;z)\tilde\chi^{-1}(- x;z)\gamma( x)
\Upsilon_0(- x;z)
\ .
}
Then, assuming that the currents $\gamma^{\mp1}\partial_\pm\gamma^{\pm1}$ and  fields $\psi_\pm$ 
fall off sufficiently fast at infinity, and remembering that $\Phi$ and  $\tilde\Phi$ depend locally on them, we have
\EQ{
\lim_{x\to\pm\infty}\Phi(x;z)=1\ ,\qquad
\lim_{x\to\pm\infty}\tilde\Phi(x;z)=1\ ,
}
and so
\EQ{
\lim_{x\to\pm\infty}\chi(x;z)=\lim_{x\to\pm\infty}\Omega(x;z)\ ,\qquad\lim_{x\to\pm\infty}\tilde\chi(x;z)=\lim_{x\to\pm\infty}\tilde\Omega(x;z)\ .
}
In addition, since $\Omega,\tilde\Omega\in\exp{\mathfrak f}^\perp$ and $\gamma(\pm\infty)\in\exp{\mathfrak f}^\perp$, this means that $\chi(\pm\infty;z)$ and $\tilde\chi(\pm\infty;z)\in\exp{\mathfrak f}^\perp$ and commute with $\Upsilon_0(x;z)$, so the subtracted monodromy 
is finally given by the two expressions:
\AL{
{\cal M}(z)&=\chi(\infty;z)\chi^{-1}(-\infty;z)=
\text{Pexp}\,\left[-\int_{-\infty}^{+\infty}dx\, \big(h_1(x;z)-z^{-2}\Lambda\big)\right]
\label{ExpInftyB}\\[5pt]
&=\gamma^{-1}(\infty)\tilde\chi(\infty;z)
\tilde\chi^{-1}(-\infty;z)\gamma(-\infty)\notag\\
&=\gamma^{-1}(\infty)\> \text{Pexp}\,\left[-\int_{-\infty}^{+\infty}dx\, \big(\tilde h_1(x;z)+z^2\Lambda\big)\right]\gamma(-\infty)\>.\label{ExpZeroB}
}
Expanding \eqref{ExpZeroB} around $z=0$ as in \eqref{mex} gives directly \eqref{chgs2}
which is the kink charge of a configuration.\footnote{Notice that in the case where $H$ is non-trivial, the  kink charge is not a true topological charge because it is not quantized at the classical level, although in the quantum theory it will, indeed, turn out to be quantized. In the case where $H=\emptyset$ the charge is a topological charge like in sine-Gordon equation.}  Notice that this is precisely the physical charge associated to global gauge transformations as argued at the end of Section \ref{sssg}.

\noindent
{\bf The Integrable hierarchy}

The integrable structure we have established above is just part of an infinite integrable hierarchy of equations. In fact, this hierarchy is an example of the generalized Drinfel'd-Sokolov hierarchies constructed in refs.~\cite{DeGroot:1991ca,Burroughs:1991bd}.\footnote{In the following, we will only need the general construction of \cite{DeGroot:1991ca,Burroughs:1991bd} with a single gradation rather than the more general possibility involving a pair of gradations. This identifies the hierarchy as being of mKdV type. Moreover,
in those references the hierarchies were classified as Type~I or~II according to whether $\hat{\mathfrak f}^\perp$ is abelian or not. For present purposes, $\hat{\mathfrak f}^\perp$ is non-abelian in general and, thus, the hierarcies will be generically of Type~II.}
Usually, when discussing an integrable hierarchy one is interested in the infinite set of flows which mutually commute. However, in the present context we are interested in flows which only commute with the spacetime flows $\partial_\pm$ but not necessarily amongst themselves. In fact, one can associate a ``flow" to each element of $\hat{\mathfrak f}^\perp$~\cite{Madsen:1999ta,Schmidtt:2009ge,Schmidtt:2010bi}, but only those associated to the (bosonic) centre of $\hat{\mathfrak f}^\perp$ will commute among themselves. 
These non-abelian flows act as hidden non-abelian symmetries of the equations-of-motion of the SSSSG theory that are not generally manifested as Noether symmetries of the action, although an important part of our argument is that a finite subset of them are.
The structure of these additional flows can be simply deduced by noticing that the Lax operators of the basic spacetime flows can be written in two distinct ways \eqref{MiniLax1} and \eqref{MiniLax2},
\EQ{
{\cal L}_\pm=\chi\big(\partial_\pm-z^{\pm2}\Lambda\big)\chi^{-1}=\gamma^{-1}\tilde{\cal L}_\pm\gamma=
\gamma^{-1}\tilde\chi\big(\partial_\pm-z^{\pm2}\Lambda\big)\tilde\chi^{-1}\gamma
}
with $\chi\in\exp\hat{\mathfrak f}_{<0}$ and $\tilde\chi\in\exp\hat{\mathfrak f}_{>0}$. Writing the basic ``spacetime flow" Lax operators in these two ways is key as we shall see.
It follows that one can construct flows which manifestly commute with them by taking
any constant element $b\in\hat f^\perp$ and defining a Lax operator
\EQ{
{\cal L}_b=\delta_b-q_b-b
=\chi\big(\delta_b-b)\chi^{-1}=
\gamma^{-1}\tilde\chi\big(\delta_b-b\big)\tilde\chi^{-1}\gamma\ .
\label{nlax}
}
If $b$ has grade $n$, then it follows from the two ways of writing ${\cal L}_b$ that $q_b$ has grades between 0 and $n-1$, for $n>0$, and between $-|n|$ and $-1$, for $n<0$.
Clearly,
\EQ{
[{\cal L}_\mu,{\cal L}_b]=0
}
and we can think of
\EQ{
\delta_b{\cal L}_\mu=[q_b + b,{\cal L}_\mu]
\label{vars}
}
as a set of symmetry variations of the original equations that are preserved under either $x^+$ or $x^-$ evolution. The consistency of \eqref{vars} requires that the right-hand side has grades $(0,1)$, for ${\cal L}_+$, and $(-2,-1)$, for ${\cal L}_-$, and this follows simply from the two ways of writing $q_b$ implicit in \eqref{nlax}. In this construction, the spacetime flows $\partial_\pm$ coincide with the flows associated to $b=z^{\pm2}\Lambda$. 
Moreover, for $n=0$ we have $q_b=0$ and $\delta_b$ corresponds just to an infinitesimal global gauge transformation.

It is important to notice that, as a direct consequence of~\eqref{nlax}, the symmetry variations form a non-abelian algebra which is simply isomorphic to $\hat{\mathfrak f}^\perp$:
\EQ{
[\delta_b,\delta_{b'}]=\delta_{[b,b']}\ .
\label{FlowAlg}
}

\noindent
{\bf Supersymmetry}

We will be particularly interested in the symmetry variations generated by the Grassmann odd elements $b=z^{\pm1}\varepsilon_\pm\in\hat{\mathfrak f}_{\pm1}$. We will argue that these variations are the odd elements of an extended SUSY algebra which are Noether symmetries of the SSSSG theory. The supersymmetry algebra corresponds to a finite subalgebra ${\mathfrak s}\subset\hat{\mathfrak f}^\perp$ generated by the elements of 
$\hat{\mathfrak f}^\perp$ of grade 0 and $\pm1$ along with two central elements of grade $\pm2$. 

To be precise, notice that the superalgebra ${\mathfrak f}^\perp\subset \mathfrak{psu}(N,N|2N)$ splits into two mutually commuting subalgebras ${\mathfrak f}^{\perp (\pm)}$ whose elements are of form
\SP{
M = \left(\begin{array}{cc|cc}m^{(+)} & 0 &-\alpha^{(+)\dagger} & 0 \\0 & m^{(-)} & 0 & \alpha^{(-)} \\\hline
\alpha^{(+)} & 0 & n^{(+)} & 0 \\0 & +\alpha^{(-)\,\dagger} & 0 & n^{(-)} \end{array}\right)\,,
\label{Form1}
}
with
\SP{
m^{(\pm)\,\dagger} = - m^{(\pm)}\,,\quad n^{(\pm)\,\dagger} = - n^{(\pm)}\,,\quad
\Tr\,\big(m^{(+)}+m^{(-)}\big)=\Tr\,\big(n^{(+)}+n^{(-)}\big)\,,
}
where all the entries are $N\times N$ matrices.
For $N=1,2$, it is easy to check that $[{\mathfrak f}^{\perp(\pm)}_1,{\mathfrak f}^{\perp(\pm)}_1]$ is a subset of ${\mathfrak f}^{\perp(\pm)}_2$ that includes only the identity element of $\mathfrak{f}^{\perp (+)}$ and $\mathfrak{f}^{\perp (-)}$, respectively; namely,
\EQ{
{\mathbb I}^{(+)}=\left(\begin{array}{cc|cc}{\mathbb I}_N &&&\\ &0&&\\ \hline &&{\mathbb I}_N&\\ &&&0\end{array}\right)\ ,
\qquad {\mathbb I}^{(-)}=\left(\begin{array}{cc|cc}0&&&\\ &{\mathbb I}_N &&\\ \hline &&0&\\  &&&{\mathbb I}_N\end{array}\right)\ .
}
However, in $\mathfrak{psu}(N,N|2N)$ one identifies elements differing by a multiple of the identity matrix and, thus, ${\mathbb I}^{(\pm)}\thicksim\mp i\Lambda$. Therefore, in $\mathfrak{psu}(1,1|2)$ and $\mathfrak{psu}(2,2|4)$, $[{\mathfrak f}^{\perp(\pm)}_1,{\mathfrak f}^{\perp(\pm)}_1]$ is actually a subset of ${\mathfrak f}^{\perp}_2$ generated by $\Lambda$, and for 
$\epsilon^{(\pm)}, \eta^{(\pm)}\in {\mathfrak f}_1^{\perp(\pm)}$ we have
\SP{
[\epsilon^{(\pm)}, \eta^{(\pm)}] = -\frac{2}{N} \Tr\big( \Lambda \epsilon^{(\pm)}\eta^{(\pm)}\big)\, \Lambda\,,
\label{Closure}
}
and a similar equation for $\epsilon^{(\pm)}, \eta^{(\pm)}\in {\mathfrak f}_{-1}^{\perp(\pm)}$.
In addition, one can check that the
subalgebra of $\mathfrak{f}^\perp$ generated by the (odd) generators in ${\mathfrak f}_{\pm1}^{\perp(\pm)}$ is just the ``derived'' algebra of ${\mathfrak f}^\perp$,
\SP{
[{\mathfrak f}^\perp,{\mathfrak f}^\perp]\subset {\mathfrak f}^\perp\,,
}
whose elements are of the form~\eqref{Form1} constrained by the additional conditions
\SP{
\Tr \big(m^{(\pm)}) = \Tr \big(n^{(\pm)}) \,.
}
All this shows that, for $N=1,2$, the subalgebra $[{\mathfrak f}^\perp,{\mathfrak f}^\perp]$ is just
\EQ{
\mathfrak{p}\big(\mathfrak{su}(N|N)\oplus \mathfrak{su}(N|N)\big)\,,
\label{loa}
}
where each of the $\mathfrak{su}(N|N)$ factors is associated to the two mutually commuting subalgebras ${\mathfrak f}^{\perp(\pm)}$, and $\Lambda$ provides a non-trivial center.

Remarkably, in the affine algebra $\hat {\mathfrak f}$ defined in~\eqref{NewAffine}, $[{\mathfrak f}^\perp,{\mathfrak f}^\perp]$ gives rise to a finite subalgebra $\mathfrak{s}$ whose generators are\footnote{The algebra also includes the Lorentz boost generator which corresponds to the derivation of the affine algebra $z d/dz$. We shall leave this implicit in the following.}
\SP{
\big\{z^{-2}\Lambda\big\} \oplus z^{-1}{\mathfrak f}^{\perp(\pm)}_{-1} \oplus z^{0}{\mathfrak h}^{(\pm)} \oplus z^{1}{\mathfrak f}^{\perp(\pm)}_{1} \oplus \big\{z^{2}\Lambda\big\}\,.
\label{Generators}
}
This is a closed algebra because of \eqref{Orthogonal2} and, crucially, because of~\eqref{Closure} that is only satisfied for $N=1,2$. Since the two generators $z^{\pm2}\Lambda$ give rise to the spacetime flows $\partial_\pm$, then the symmetry variations~\eqref{vars} associated to $z^{-1}{\mathfrak f}^{\perp(\pm)}_{-1}$ and $z^{+1}{\mathfrak f}^{\perp(\pm)}_{+1}$ are actually SUSY transformations that, using~\eqref{FlowAlg} and~\eqref{Closure}, satisfy
\SP{
&\big[\delta_{z^{+1}\epsilon^{(\pm)}}, \delta_{z^{+1}\eta^{(\pm)}}\big]
= -\frac{2}{N} \Tr\big( \Lambda \epsilon^{(\pm)}\eta^{(\pm)}\big)\; \partial_+\,,\qquad \epsilon^{(\pm)},\eta^{(\pm)} \in {\mathfrak f}^{\perp (\pm)}_{+1}\,,\\[5pt]
&\big[\delta_{z^{-1}\tilde\epsilon^{(\pm)}}, \delta_{z^{-1}\tilde\eta^{(\pm)}}\big]
= -\frac{2}{N} \Tr\big( \Lambda \tilde\epsilon^{(\pm)}\tilde\eta^{(\pm)}\big)\; \partial_-\,,\qquad \tilde\epsilon^{(\pm)},\tilde\eta^{(\pm)} \in {\mathfrak f}^{\perp (\pm)}_{-1}\,.
\label{salgg}
} 
Moreover, the comparison between $\mathfrak{s}$ and the form or $[\mathfrak{f}^\perp,\mathfrak{f}^\perp]$ given by~\eqref{loa} shows that the SUSY algebra $\mathfrak{s}$ is actually isomorphic to the double central extension
\EQ{
\big(\mathfrak{psu}(N|N)\oplus \mathfrak{psu}(N|N)\big)
\ltimes\big( \R\oplus\R\big)\,,
\label{loaB}
}
where the two generators of the central extension $\R\oplus\R$ correspond to $z^{-2}\Lambda$ and $z^{+2}\Lambda$ or, equivalently, to the spacetime flows $\partial_\pm$. 

Now we consider more details of the 2 cases $N=1,2$ separately.

{\bf (i)} For the more complicated $\mathfrak{psu}(2,2|4)$ case,
the complete extended SUSY algebra is $\mathfrak{psu}(2|2)\oplus \mathfrak{psu}(2|2)$ with two central extensions,  
\EQ{
{\mathfrak s}=\big(\mathfrak{psu}(2|2)\oplus\mathfrak{psu}(2|2)\big)\ltimes\big({\mathbb R}\oplus\mathbb R\big)
}
and the corresponding supergroup is 
\EQ{
{\cal S}=PSU(2|2)^{\times2}\ltimes{\mathbb R}^{\times2}\ .
\label{ssalg}
}
Each $\mathfrak{psu}(2|2)$ factor provides 4~SUSY generators of each chirality (4 in $z^{1}\otimes \mathfrak{f}_{+1}^{\perp(\pm)}$ and 4 in $z^{-1}\otimes \mathfrak{f}_{-1}^{\perp(\pm)}$). Therefore, there are 8 independent real supersymmetries in each $PSU(2|2)$ factor and so the SUSY is an exotic example of ${\cal N}=(8,8)$. Moreover, the bosonic subgroup includes the group of global gauge transformations $H=SU(2)^{\times4}$. The novel thing is that H plays the r\^ole of what appears to be a non-abelian R-symmetry group, with the supercharges transforming under $H$ and the physical bosonic and fermionic degrees-of-freedom come in different representations of this group; that is $\mathfrak f_2^\parallel=(2,2,1,1)+(1,1,2,2)$, for the bosons, and $\mathfrak f_1^\parallel+\mathfrak f_3^\parallel=(2,1,1,2)+(1,2,2,1)$ for the fermions. 

{\bf (ii)} For the case $\mathfrak{psu}(1,1|2)$ the situation is much simpler. The (bosonic) 0-graded subalgebras $\mathfrak{h}^{(\pm)}$ are absent, so there is no R symmetry group, and $\mathfrak{s}$
takes the form of a conventional supersymmetry algebra with 2 independent real supersymmetries. In particular, the generators in $z^{-1}\mathfrak f_{-1}^{\perp(\pm)}$ commute with those in $z\mathfrak f_{+1}^{\perp(\pm)}$. This dovetails nicely with the fact that the physical fields are abelian. The number of independent SUSYs is ${\cal N}=(2,2)$.

It is important to note that if we try to generalize the theories to arbitrary $\mathfrak{psu}(N,N|2N)$ then the algebra generated by the Grassmann elements ${\mathfrak f}^\perp_{\pm1}$ does not close onto a finite subalgebra of $\hat{\mathfrak f}^\perp$. Therefore, the resulting theories will not have a conventional supersymmetry.

The variations of the fields can be found by expanding out \eqref{vars} and solving for the variations self consistently. For $b=z\varepsilon_+$, one finds
\EQ{
\gamma^{-1}\delta\gamma&=-2\Lambda[\varepsilon_+,\psi_+]-q^\perp\ ,\\
\delta\psi_+&=[\varepsilon_+,(\gamma^{-1}\partial_+\gamma)^\parallel]+[q^\perp,\psi_+]\ ,\\
\delta\psi_-&=2\Lambda(\gamma\varepsilon_+\gamma^{-1})^\parallel\ .
\label{susv}
}
In the above, $q^\perp$ can be thought of as a compensating gauge transformation that is needed to preserve the gauge constraints \eqref{cons}. This term is non-local in the fields and is found by integrating
\EQ{
\partial_-q^\perp=[\varepsilon_+,(\gamma^{-1}\psi_-\gamma)^\perp]=[\varepsilon_+,{\cal G}^{(+)}_-]
\label{eqd}
}
This complicating factor is not present in the simpler $\mathfrak{psu}(1,1|2)$ theories.
It is important that these variations preserve the gauge conditions \eqref{cons} and the kappa symmetry conditions \eqref{kapf}. There is a similar set of variations associated to $z^{-1}\varepsilon_-$ that can be found by making the substitutions \eqref{Parity}. 

The supersymmetry transformations have been constructed on-shell, so are symmetries of the equations-of-motion in the on-shell gauge $A_\mu=0$. However, in the Appendix
we show that these symmetries extend off-shell to symmetries of the action.  For the $\mathfrak{psu}(2,2|4)$ theory the non-trivial aspect of these transformations is that they involve the non-local gauge transformation $q^\perp$ which explains why they are not obvious symmetries of the action. Ultimately it seems likely that this non-local nature of the transformations could be responsible in the quantum theory for a $q$ deformation of the supersymmetry algebra $\mathfrak s$. 
For the simpler $\mathfrak{psu}(1,1|2)$ theories, the supersymmetry has been identified in  \cite{Grigoriev:2007bu,Grigoriev:2008jq} as conventional ${\cal N}=2$ supersymmetry.

\section{The Perturbative Spectrum}
\label{pmass}

In this section we establish the spectrum of perturbative fluctuations. Taking $A_\mu=0$, the linearized equations-of-motion for fluctuations $\phi$ where
\EQ{
\gamma=e^\phi \simeq 1+\phi+\cdots
}
is simply the free wave equation
\EQ{
\square\phi=\big(\partial_0^2-\partial_1^2)\phi =4\big[\Lambda,[\Lambda,\phi]\big]\ .
}
The on-shell gauge conditions \eqref{cons} have the effect of removing the massless modes $\phi^\perp\in{\mathfrak h}$. In order to see this, expand 
\EQ{
\phi=\phi^\perp+\phi^\parallel
}
and solve the constraints \eqref{cons} for $\phi^\perp$ order-by-order in the fluctuation $\phi^\parallel$. To lowest order
\EQ{
\partial_\pm\phi^\perp=\pm\frac12[\phi^\parallel,\partial_\pm\phi^\parallel]\mp2\Lambda\psi_\pm\psi_\pm+\cdots\ .
\label{wer}
}
In the above, $\psi_\pm\equiv\psi_\pm^\parallel$ due to the constraint \eqref{kapf}.
Hence, to linear order $\phi^\perp=0$, however, it is interesting that to quadratic order
$\phi^\perp$ becomes non-vanishing. Pursuing this further we find that $\phi^\perp$ actually has a kink-like behaviour; to quadratic order
\EQ{
&\phi^\perp(x=\infty)-\phi^\perp(x=-\infty)\\ &=\int_{-\infty}^\infty dx\, \partial_1
\phi^\perp=
\int_{-\infty}^\infty dx\,\Big(\frac12
[\phi^\parallel,\partial_0\phi^\parallel]+2\Lambda\big(\psi_+\psi_++\psi_-\psi_-\big)\Big)\ .
}
Remarkably, the right-hand side is the $H$ charge of a perturbative mode. In order to see this, note that the tree-level action for the perturbative modes is
\EQ{
S=-\frac k{\pi}\int d^2x\,&\Tr\Big(\frac18\partial_\mu\phi^\parallel\partial^\mu\phi^\parallel-\frac12[\Lambda,\phi^\parallel]^2\\ 
&+\Lambda\psi_+\partial_-\psi_+-\Lambda\psi_-\partial_+\psi_-+\psi_+\psi_-+\cdots\Big)\ ,
}
which is invariant under adjoint action under the unbroken global part of the gauge symmetry:
\EQ{
\phi^\parallel\longrightarrow U\phi^\parallel U^{-1}\ ,\qquad \psi_\pm\to U\psi_\pm U^{-1}\ .
}
The associated Noether current takes the form
\EQ{
{\cal J}_\pm=\pm\frac12[\phi^\parallel,\partial_\pm\phi^\parallel]\pm2\Lambda\psi_\pm\psi_\pm\ .
\label{ncur}
}
Consequently, as anticipated in \eqref{chgs1} and \eqref{chgs2}, the Noether charge is equal to the kink charge
\EQ{
{\cal Q}=\int_{-\infty}^\infty dx\,{\cal J}^0=-\phi(\infty)+\phi(\infty)\equiv q_0\ .
}

In fact the $H$ symmetry of the 
free action for the perturbative modes is actually part of a larger invariance under the extended supersymmetry algebra $\mathfrak s$ whose variations follow by expanding  \eqref{susv} to linear order:
\EQ{
\delta\phi^\parallel=-2\Lambda[\varepsilon_+,\psi_+]\ ,\qquad
\delta\psi_+=[\varepsilon_+,\partial_+\phi^\parallel]\ ,\qquad
\delta\psi_-=2\Lambda[\phi^\parallel,\varepsilon_+]\ .
\label{susvLin}
}
We can now go on to quantize the perturbative modes at tree level
and this will lead to a quantization of the kink charge. The perturbative states naturally fall into the fundamental representation of the 
supergroup ${\cal S}$, that is the $(2|2)$-dimensional representation for each of the $SU(2|2)$ factors, for the $\mathfrak{psu}(2,2|4)$ case.

\section{The Solitons}

In this section we construct soliton solutions. The discussion will mostly be aimed to the more involved $AdS_5\times S^5$ case, but can easily be adapted to the simpler $AdS_2\times S^1$ case. The idea will be to generalize the dressing method used in the bosonic theories and described in detail in \cite{Hollowood:2010dt}. The soliton solutions will be solutions of the equations-of-motion with
$A_\mu=0$, and which automatically satisfy the gauge constraint \eqref{cons} and the kappa symmetry conditions \eqref{kapf}.

The dressing method focuses on the solution of the linear system
\eqref{LinProb}.
The quantity $\Upsilon(z)$ is an element of the loop group associated to $\hat{\mathfrak f}$ and so it must satisfy the reality condition \eqref{real} lifted to the loop group
\EQ{
H{\Upsilon(z^*)^{-1}}^\dagger H=\Upsilon(z)\ .
\label{cond2}
}
Similarly it must have the appropriate behaviour under the automorphism
\EQ{
{\cal K}^{-1}{\Upsilon(z)^{-1}}^{st}{\cal K}=\Upsilon(iz)\ .
\label{cond1}
}
It is also useful to have the action of fermionic parity
\EQ{
{\mathfrak P}\Upsilon(z){\mathfrak P}=\Upsilon(-z)\ .
\label{cond3}
}

Soliton solutions are special solutions 
for which $g_+=g_-=1$ in the Riemann-Hilbert problem \eqref{RHp} 
~\cite{Babelon}. Then,~\eqref{LPsol1} and~\eqref{LPsol2} imply that the solution of the linear problem can be written in two equivalent ways
\EQ{
\Upsilon(x;z)=\chi(x;z)\Upsilon_0(x;z)=\gamma^{-1}\tilde\chi(x;z)\Upsilon_0(x;z)\,.
\label{dres}
}
In the context of solitons, $\chi(z)\equiv\chi(x;z)$ is known as the ``dressing transformation" for the obvious reason that it generates the soliton solutions from the vacuum. The method then proceeds by taking an ansatz for the dressing factor which takes the form of a sum over a finite set of simple poles
\EQ{
\chi(z)=1+\frac{Q_i}{z-\xi_i}\ ,\qquad
\chi(z)^{-1}=1+\frac{R_i}{z-\mu_i}\ .
}
Then, the associated linear problem~\eqref{LinProb} (in the gauge $A_\pm=0$) gives rise to the two equations:
\AL{
&\partial_+\chi(z) \chi(z)^{-1} + z^2\chi(z) \Lambda\chi(z)^{-1} = -
\gamma^{-1}\partial_+\gamma-z\psi_++z^2\Lambda\ ,
\label{Eq1}\\[5pt]
&\partial_-\chi(z) \chi(z)^{-1} 
+ z^{-2}\chi(z) \Lambda\chi(z)^{-1} =-z^{-1}\gamma^{-1}\psi_-\gamma+ z^{-2}\gamma^{-1}\Lambda\gamma\ .
\label{Eq2}
}

The fields can be extracted from the expansions of $\chi(z)$ around $z=0$ and $z=\infty$:
\EQ{
\chi(z)&=1+z^{-1}W_{-1}+z^{-2}\big(W_{-2}+\tfrac12W_{-1}^2\big)
+{\cal O}(z^{-3})\\
&=\gamma^{-1}\Big(1+zW_1+z^2\big( W_2+\tfrac12 W_1^2\big)+
{\cal O}(z^3)\Big)\ .
}
Hence, as well as
\EQ{
\gamma=\chi(0)^{-1}\ ,
}
we have
\EQ{
\psi_\pm=[W_{\mp1},\Lambda]\ ,
}
which is the kappa symmetry fixing condition \eqref{kapf}.
At the next order $z^0$, we find
\EQ{
\gamma^{\mp1}\partial_\pm\gamma^{\pm1}+2\Lambda\psi_\pm\psi_\pm
=[\Lambda,W_{\mp2}] +\frac{1}{2}[W_{\mp1}^\perp,\psi_\pm]\ ,
}
which implies that the projection of the left-hand side into ${\mathfrak h}$ vanishes. These are precisely the on-shell gauge constraints \eqref{cons}. 

Now, since the dependence on $z$ of the right-hand-side of \eqref{Eq2}
is explicit, 
the residues of the left-hand-side at $z=\xi_i$ and $\mu_i$ must vanish, giving
\SP{
\left(\xi_i^{\mp2} \partial_\pm Q_i + Q_i
  \Lambda\right)\Big(1+\frac{R_j}{\xi_i-\mu_j}\Big) &=0\ ,\\[5pt]
\Big(1+\frac{Q_j}{\mu_i-\xi_j}\Big)\left(-\mu_i^{\mp2}\partial_\pm
  R_i + \Lambda R_i\right)&=0\ .
\label{kss}
}
The key  
to solving them is to propose that the residues have rank one~\cite{Harnad:1983we,Hollowood:2010dt}:
\EQ{
Q_i = \BX_i \BF_i^\dagger \quad\text{and}\quad
R_i = \BH_i \BK_i^\dagger\ ,
}
where 8-vectors are written in boldface. However, at this point we have to make a choice. The point is that in order to preserve the fermionic grading, the vectors must have the 
structure
\EQ{
\Bv=\left(\begin{array}{c} \Bv_1 \\ \hline \Bv_2\end{array}\right)
\label{dpp}
}
where either of the 4-vectors $\Bv_1$, or $\Bv_2$, must be Grassmann odd. This ensures that a matrix of the form $\Bv\Bw^\dagger$ is valued in $\mathfrak{gl}(4|4)$.
We shall fix the choice by realizing that there is a known consistent bosonic soliton solution for the $S^5=SU(4)/Sp(4)$ factor. This solution would be obtained by taking a dressing ansatz where all the vectors have $\Bv_1=0$. In other words, the sub 4-vector $\Bv_1$ must be Grassmann odd and the sub 4-vector $\Bv_2$ must be Grassmann even.
The other possible solutions with $\Bv_1$ Grassmann even should be in the domain of spiky strings configurations, which are soliton-like solutions with singularities at the spikes.

Notice that the fermionic parity operator ${\mathfrak P}$ has the correct action on these vectors which we call $(4|4)$ vectors. These vectors have the following properties: for 2 such vectors we have
\SP{
&
\big(\Bv\Bw^\dagger\big)^\dagger=\Bw\Bv^\dagger\ ,\qquad
\big(\Bv\Bw^t\big)^{st}={\mathfrak P}\Bw\Bv^t\ ,\qquad\Bv\cdot\Bw=\Bw\cdot{\mathfrak P}\Bv\ ,\\[5pt]
&
\big(\Bv^*\cdot\Bw)^*=\Bw^*\cdot\Bv\ ,\qquad\text{STr}\,(\Bv\Bw^\dagger)=\Bw^*\cdot\Bv\ .
}

The solution of \eqref{kss} is
\EQ{
\BF_i=\big(\Psi_0(\xi_i)^\dagger\big)^{-1}\Bvarpi_i\ ,\qquad
\BH_i=\Psi_0(\mu_i)\Bpi_i\ ,
}
for constant complex graded $8$-vectors $\Bvarpi_i$ and $\Bpi_i$ along with
\EQ{
\BX_i\Gamma_{ij}=\BH_j\ ,\qquad \BK_i(\Gamma^\dagger)_{ij}=-\BF_j\ ,
}
where the matrix
\EQ{
\Gamma_{ij}=\frac{\BF_i^*\cdot \BH_j}{\xi_i-\mu_j}\ .
}

At the moment, we have a ``raw" solution of the linear system but this needs to be refined so that it satisfies the conditions \eqref{cond2}, \eqref{cond1} and \eqref{cond3}. 
Implementing the reality condition \eqref{cond2} gives
\EQ{
\frac{H\big(\BH_i\BK_i^\dagger\big)^\dagger H}{z-\mu_i^*}=\frac{\BX_j\BF_j^\dagger}{z-\xi_j},
}
which is solved by taking
\EQ{
\mu_i=\xi_i^*,\qquad
\BK_i=H\BX_i,\qquad
\BH_i=H\BF_i,
}
and so
\EQ{
\Gamma_{ij}=\frac{\BF_i^*\cdot H\BF_j}{\xi_i-\xi_j^*} = - \Gamma_{ji}^*\ .
}

Similarly, the condition \eqref{cond1} gives
\EQ{
\frac{{\cal K}^{-1} \big(\BH_i\BK_i^\dagger\big)^{st}{\cal K}}{z-\xi_i^*}=\frac{\BX_j\BF_j^\dagger}{iz-\xi_j},
}
which means that as a set $\{\xi_i^*\}=\{-i\xi_i\}$. Consequently, we define the operator $\eta$ with
\EQ{
\xi_i^*=-i\xi_{\eta(i)}\ ,\qquad
\BF_i=\varepsilon_i H{\cal K}\BF_{\eta(i)}^*\ ,\qquad 
\BX_i= i \varepsilon_i  {\mathfrak P}H{\cal K} \BX_{\eta(i)}^*\ ,
\label{SignsA}
}
where $\varepsilon_i=\pm1$. It is important to notice that $\eta^2(i)=i$:
\EQ{
\xi_{\eta^2(i)} = i \xi_{\eta(i)}^* = i \big(i\xi_i^*\big)^*=\xi_i,
}
which constraints the choice of the signs $\varepsilon_i$:
\EQ{
\BF_i=\varepsilon_i H{\cal K}\BF_{\eta(i)}^* =\varepsilon_i H{\cal K} \big(\varepsilon_{\eta(i)}H{\cal K}\BF_{\eta^2(i)}^* \big)^* =-\varepsilon_i\varepsilon_{\eta(i)}\BF_i\ ,
}
where we have used that ${\cal K}^2=-1$. Therefore, we have the constraint
\EQ{
\varepsilon_i\varepsilon_{\eta(i)}=-1\ ,
}
and we shall choose
\EQ{
\varepsilon_1=\varepsilon_3=-1\ ,\qquad\varepsilon_2=\varepsilon_4=1\ .
}

Finally, the condition \eqref{cond3} gives
\EQ{
\frac{{\mathfrak P} \BX_i\BF_i^\dagger{\mathfrak P}}{z-\xi_i}=-\frac{\BX_j\BF_j^\dagger}{z+\xi_j}.
}
Therefore, as a set $\{\xi_i\}=\{-\xi_i\}$, and
\EQ{
\xi_i=-\xi_{\rho(i)} \;\Rightarrow\; \BX_i= -{\mathfrak P}\BX_{\rho(i)},\qquad \BF_i= {\mathfrak P}\BF_{\rho(i)}\ ,
\label{SignsB}
}
with $\rho^2(i)=i$.

Taken together, these conditions require the ``dressing data" to have four poles. Choosing the ordering
\EQ{
\{\xi_i\}= \{\xi, i\xi^*, -\xi, -i\xi^* \}, \qquad
\label{has}
}
we have $\eta(1,2,3,4)=(2,1,4,3)$ and $\rho(1,2,3,4)=(3,4,1,2)$ and
\EQ{
\{\BF_i\}= \{\BF, {\cal K}H\BF^*, {\mathfrak P}\BF, {\cal K}H{\mathfrak P}\BF^* \}\ , 
\label{hat}
}
which means that the constant $(4|4)$ vectors are
\EQ{
\{\Bvarpi_i\}=\{\Bvarpi, {\cal K}H\Bvarpi^*, {\mathfrak P}\Bvarpi, {\cal K}H{\mathfrak P}\Bvarpi^* \}\ ,
}

For later use, we define the operators $\sigma_i$, $i=1,2,3,4$, such that $\sigma_i(\xi)=\xi_i$ and $\sigma_i(\Bvarpi)=\Bvarpi_i$, with $\sigma_1(\xi)\equiv\xi$ and $\sigma_1(\Bvarpi)\equiv\Bvarpi$. Using this notation, the dressing factor is
\EQ{
\chi(x;z)=1+\frac{H\sigma_i(\BF){\Gamma_{ij}^{-1}}\sigma_j(\BF)^\dagger}
{z-\sigma_j(\xi)}\ .
}

\noindent
{\bf Collective coordinates and mass}

Since the $x$ and $t$ dependence of the soliton are encoded in 
$\Upsilon_0(\xi)$, writing
$\xi=e^{-\vartheta/2-iq/2}$ identifies $\vartheta$ with the rapidity. 
It is also useful to write 
\EQ{
\Bvarpi=\Bv_++\Bv_-\ ,
}
where $\Bv_\pm$ are eigenvectors of $\Lambda$ of eigenvalue $\mp i/2$ in the degenerate subspaces. We will take
\EQ{
\Bv_+=\MAT{0\\ \Bu_+\\  
\hline 0\\ \BOmega_+}\,,\qquad 
\Bv_-=\MAT{\Bu_-\\ 0\\ \hline \BOmega_- \\ 0}\ ,\qquad\ .
}
In the soliton rest frame, which corresponds to $\vartheta=0$,
\EQ{
\BF=\exp\big(it\cos q+x\sin q\big)\Bv_++\exp\big(-it\cos q-x\sin q\big)\Bv_-\ .
}
Physically inequivalent solutions are obtained by restricting $0\leq q\leq\frac\pi2$.
Consequently, since $\sin q>0$, in the asymptotic regimes $x\to\pm \infty$ we can effectively replace 
$\Bvarpi$ by $\Bv_\pm$, respectively. Since the solution is invariant under the complex re-scalings $\Bvarpi\to\lambda\Bvarpi$ this means that it becomes independent of $x$ and $t$ in the asymptotic regimes which is consistent with it being a localized soliton. 

The freedom in the vectors $\Bv_\pm$  implies that
the kink carries internal collective coordinates. In order to understand their significance, consider the action of the SUSY group ${\cal S}$ on the solution.
This can be obtained by considering an ${\cal S}$ variation of the vacuum solution and then by dressing this in the standard way to construct the transformed soliton. The transformation of the vacuum under a symmetry $b\in\mathfrak s$ is simply given by 
\EQ{
\big(\delta_b-b\big)\Upsilon_0(z)=0\ ,\qquad b\in\mathfrak{s}
}
and so the transformed vacuum solution is in the orbit of the supergroup \eqref{ssalg} acting in the rest frame of the soliton which we denote ${\cal S}_0$. In the rest frame, the 
central extensions are equal, {\it i.e.\/}~$z=1$ in~\eqref{Generators}, and so
\EQ{
{\cal S}_0=P\big(SU(2|2)^{\times2}\big)\ .
}
Therefore the orbit is of the form
\EQ{
\Upsilon_0(x;z)\longrightarrow\Upsilon_0(x;z){\cal U}\ ,
}
where ${\cal U}$ is a constant ($x^\pm$ independent) element of ${\cal S}_0$. 
Notice that ${\cal U}$ includes SUSY transformations and global gauge transformations. When the transformed vacuum is dressed, it gives rise to a soliton solution with a transformation of $\Bvarpi$:
\EQ{
\Bvarpi\longrightarrow{\cal U}\Bvarpi\ .
}
This freedom, along with scaling symmetries and spacetime translations, 
can be used to set
\EQ{
\BOmega_\pm=\MAT{0\\ 1} \ ,\qquad\Bu_\pm=\MAT{0\\ 0}\ .
\label{spc}
}
which is the bosonic kink solution with the fermionic fields vanishing. Hence, just as in the SSSG theories, the kinks have an internal moduli space that can be thought of as a (co-)adjoint orbit, but in this case of the supergroup ${\cal S}_0$. The (co-)adjoint orbit in question takes the form of a product of two cosets of the form
\EQ{
\CP^{2|1}=\frac{SU(2|2)}{U(2|1)}\ .
}
where $U(2|1)$ is the stability group of \eqref{spc}.

The conserved charges can be calculated from the subtracted monodromy \eqref{ExpInftyB}
\EQ{
{\cal M}(z)&=\chi(x=\infty;z)\chi(x=-\infty;z)^{-1}\\
&=\Big(1+\frac{H\sigma_i(\Bv_{+}){\Gamma_{ij}^{(+)-1}}\sigma_j(\Bv_{+})^\dagger}
{z-\sigma_j(\xi)}
\Big)\Big(1-\frac{H\sigma_k(\Bv_-){\Gamma_{kl}^{(-)-1}}\sigma_l(\Bv_-)^\dagger}
{z-\sigma_k(\xi^*)}\Big)\\
&=1+\frac{H\sigma_i(\Bv_{+}){\Gamma_{ij}^{(+)-1}}\sigma_j(\Bv_{+})^\dagger}
{z-\sigma_j(\xi)}
-\frac{H\sigma_k(\Bv_-){\Gamma_{kl}^{(-)-1}}\sigma_l(\Bv_-)^\dagger}
{z-\sigma_k(\xi^*)}\ ,
\label{kcha}
}
using $\sigma_i(\Bv_+^*)\cdot H\sigma_j(\Bv_-)=0$,
where we have defined
\EQ{
\Gamma^{(\pm)}_{ij}=\Gamma_{ij}\Big|_{\BF\to \Bv_\pm}=
\frac{\sigma_i(\Bv_\pm)^*\cdot H \sigma_j(\Bv_\pm)}{\sigma_i(\xi)-\sigma_j(\xi^*)}\ .
}
In the following, it will also be useful to define
\EQ{
[G^{(\pm)}_{m,n}]_{ij}=\sigma_i(\xi^*)^m\Big(\sigma_i(\Bv_\pm)^*\cdot H \sigma_j(\Bv_\pm)\Big)\sigma_j(\xi)^n\ .
}

Let us extract the mass of the soliton. The observations
\EQ{
\text{STr}\,(\Lambda q_{-1}^2)&=\frac i2\,\Tr\Big[({\Gamma^{(+)}}^{-1}G_{0,0}^{(+)})^2-
({\Gamma^{(-)}}^{-1}G_{0,0}^{(-)})^2\Big]=0\ ,\\
\text{STr}\,(\Lambda q_{1}^2)&=\frac i2\,\Tr\Big[\big({\Gamma^{(+)}}^{-1}G^{(+)}_{-2,0}+
{\Gamma^{(+)}}^{-1}G^{(+)}_{-2,-1}{\Gamma^{(+)}}^{-1}G_{0,0}^{(+)}\big)^2\\ &
-\big({\Gamma^{(-)}}^{-1}G^{(-)}_{0,-2}+
{\Gamma^{(-)}}^{-1}G^{(-)}_{-1,-2}{\Gamma^{(-)}}^{-1}G_{0,0}^{(-)}\big)^2\Big]=0
}
mean that
\EQ{
p_\pm=\mp\frac k{2\pi}\text{STr}\,(\Lambda q_{\mp2})&=\mp\frac k{2\pi}\text{STr}\,\left[\Lambda{\cal M}(z)\right]_{z^{\mp2}}\ .
}
The notation here means picking the coefficient of $z^{-2}$ or $z^2$ in the expansion around $\infty$ and $0$, respectively. Hence, using \eqref{kcha},
\EQ{
p_+
&=-\frac{ik}{4\pi}\Tr\Big[{\Gamma^{(+)}}^{-1}G^{(+)}_{1,0}+{\Gamma^{(-)}}^{-1}G^{(-)}_{0,1}\Big]\\
&=\frac {ik}{2\pi}(\xi^{*2}-\xi^2)=\frac k\pi e^{-\vartheta}\sin q
}
and 
\EQ{
p_-
&=\frac{ik}{4\pi}\Tr\Big[-{\Gamma^{(+)}}^{-1}G^{(+)}_{-3,0}-{\Gamma^{(+)}}^{-1}G^{(+)}_{-3,-1}-{\Gamma^{(-)}}^{-1}G^{(-)}_{0,-3}-{\Gamma^{(-)}}^{-1}G^{(-)}_{-1,-3}\Big]\\
&=\frac{ik}{2\pi}(\xi^{-2}-\xi^{*-2})=\frac k\pi e^{+\vartheta}\sin q\ .
}
Hence, the soliton has a mass\footnote{Recall that we set the overall mass scale to $\mu=1$.}
\EQ{
M=\frac{2k}\pi\sin q\ .
\label{msol}
}
This matches the mass calculated in the bosonic SSSG theories in \cite{Hollowood:2010dt}.

\section{Semi-Classical Quantization}

In the remainder of the paper, we proceed to a semi-classical quantization of the soliton in the standard way. Namely; one allows the collective coordinates to become time-dependent $X\to X(t)$ and then substitutes the solution back into the action. Integrating over the spatial domain gives an effective action for the collective coordinate functions $S_\text{eff}[X(t)]$. This one-dimensional field theory, which can be viewed as existing along the world line of the soliton, can then be quantized leading to quantum mechanics on the moduli space ${\mathfrak M}$. To lowest order in the semi-classical limit one keeps the terms with the smallest number of $t$-derivatives. In usual soliton theories these terms are quadratic in $\dot X$. However, as was shown in \cite{Hollowood:2010dt,Hollowood:2011fm}, for the SSSG theories the dominant terms are actually linear in $\dot X$, and on quantization this leads to a non-commutative version of the moduli space. The new element in the present setting are the Grassmann odd coordinates: $\mathfrak M$ is a superspace. Since the fermionic kinetic terms are linear in $t$ derivatives we suspect that the effective action of the Grassmann odd coordinates is also linear in their $t$-derivatives. This means that SUSY has a interesting realization on the effective quantum mechanics on the moduli space: it is a global symmetry of a supergroup rather than a local symmetry along the world line.

In this regard, the Grassmann even and odd collective coordinates will be treated somewhat differently. For fixed charge $q$, the bosonic soliton solution, with all the Grassmann coordinates set to zero, is a bona-fide semi-classical object that consists of a large number of quantum excitations. In fact, it was found in \cite{Hollowood:2010dt} that this picture can be made very concrete: the semi-classical soliton states are coherent states of high excitation number of order $k$. In contrast, for the Grassmann odd sector these kind of semi-classical states do not exist,  and because of Fermi statistics there are only a small number of states with low occupation number. Consequently, the Grassmann odd exictations are inherently quantum.  Although we have
the full classical solution to all orders in the Grassmann odd coordinates, in the quantum theory the interpretation of this general solution with the Grassmann odd modes turned on is potentially beset by operator ordering ambiguities. Consequently, we will work to lowest order in the Grassmann odd coordinates as this will be enough to see the supersymmetry of the quantum mechanical system. 

First of all, we consider the bosonic solution with all the Grassmann odd coordinates turned off $\Bu_\pm=0$. In this case, following the discussion in \cite{Hollowood:2010dt,Hollowood:2011fm}, the moduli space in the Grassmann even directions takes the form of a co-adjoint orbit of the subgroup $H_b=SU(2)_b^+\times SU(2)_b^-\subset H$.
In order to find the effective quantum mechanical action,  we substitute
\EQ{
\gamma\longrightarrow U(t)\gamma\, U(t)^{-1}\ ,\qquad U\in H_b\subset H
}
into the action of the theory. It is important to notice that this is not a gauge transformation since we keep $A_\mu=0$. In \cite{Hollowood:2010dt}, it was proved that the effective action for $U(t)$ is
\EQ{
S_\text{eff}[U]&=
\frac k{2\pi}\int d^2x\,\partial_\mu\Tr\,\big(U^{-1}\dot U{\cal J}^\mu\big)\\
&=\frac k{2\pi}\int dt\,\Tr\Big[U^{-1}\dot U\,\int dx\,{\cal J}^0\Big]\\ &
=\frac k{2\pi}\int dt\,\Tr\Big[U^{-1}\dot U\,q_0\Big]\ ,
}
to (first) linear order in the time derivative $U^{-1}\dot U$. For the bosonic solution
\EQ{
q_0=2iq\left(\begin{array}{cc|cc} 0 &  &  & \\
  & 0 &  & \\ \hline
 &  & \BOmega_{+}\BOmega_{+}^\dagger-J_2\BOmega_{+}^*\BOmega_{+}^tJ_2^{-1} & \\
 &  & & - 
\BOmega_{-}\BOmega_{-}^\dagger+J_2\BOmega_{-}^*\BOmega_{-}^tJ_2^{-1}\end{array}\right)\ .
\label{topc4}
}
Alternatively we can write the effective action for the collective coordinates in terms of the unit 2-vectors $\BOmega_\pm(t)=U(t)\BOmega_\pm^{(0)}$, where $\BOmega_\pm^{(0)}$ are some arbitrary reference vectors, as
\EQ{
S_\text{eff}=\frac {2ikq}{\pi}\int dt\,\left(\dot\BOmega_+^*\cdot\BOmega_+-\dot\BOmega_-^*\cdot\BOmega_-\right)\ .
\label{EffB}
}

From now on, we focus on the $+$ sector and drop the label: the $-$ sector is similar. What we have classically is a 
mechanical system with an enlarged phase space paramaterized by the complex 2-vector $\BOmega$ with Poisson brackets
\EQ{
\{\BOmega_i,\BOmega^*_j\}=\frac{i\pi}{2qk}\delta_{ij}\ .
} 
The physical phase space, which is identified with the moduli space of the soliton, 
involves a K\"ahler quotient. This starts by noticing that 
the $U(1)$ symmetry $\BOmega\to e^{i\alpha}\BOmega$ 
is a Hamiltonian symmetry generated by 
$\Phi=\BOmega^*\cdot\BOmega$. The physical phase space corresponds to restricting $\BOmega$ to the level set 
\EQ{
\Phi=\BOmega^*\cdot\BOmega=1
}
and performing a quotient by the $U(1)$ symmetry. This is the familiar K\"ahler quotient construction of  $\CP^{1}$.

In the quantum theory, we can replace the Poisson brackets by commutators 
involving the operators $\hat\BOmega_i$ and $\hat\BOmega_i^\dagger$:
\EQ{
[\hat\BOmega_i,\hat\BOmega_j^\dagger]=\frac{\pi}{2qk}\delta_{ij}
\label{nco}
}
and build a Hilbert space by treating the former as annihilation operators and the latter as creation operators if $q>0$. If $q<0$ the r\^oles of the operators are interchanged. 
The Hamiltonian 
\EQ{
\hat\Phi=\frac\pi{2qk}\hat\BOmega^\dagger\cdot\hat\BOmega
} 
is proportional to the number operator and the constraint is just the condition that the occupation number is 
\EQ{
{\EuScript N}=\frac{2qk}{\pi}=1,2,\ldots\ .
}
This involves a quantization of $q$ 
\EQ{
q=\frac{\pi {\EuScript N}}{2k}\ ,\qquad {\EuScript N}=1,2,\ldots,k\ ,
\label{iuu}
}
where we have taken account of the fact that $q$ is restricted to lie in the range $0\leq q\leq\tfrac\pi2$. 
Then, the Hilbert space is spanned by the states\footnote{Notice that the quotient by $U(1)$ is trivial at the level of the Hilbert space.}
\EQ{
\hat\BOmega_{i_1}^\dagger\hat\BOmega_{i_2}^\dagger\cdots\hat\BOmega_{i_{\EuScript N}}^\dagger|0\rangle\ ,
}
which identifies it as the representation space for the spin $\frac{\EuScript N}2$ representation of $SU(2)$. Now we take account of the fact that there are two $\pm$ sectors which means that the soliton states are actually  the spin $(\frac{\EuScript N}2,\frac{\EuScript N}2)$ states of the product $SU(2)_b^{(+)}\times SU(2)_b^{(-)}$.

The quantization we have found means that the masses of the solitons are discrete:
\EQ{
M=\frac{4k}\pi\sin\left(\frac{\pi{\EuScript N}}{2k}\right)\ ,\qquad{\EuScript N}=1,2,\ldots,k\ .
}
 
\noindent{\bf The Grassmann odd coordinates} 
 
Now we must consider the effects of the Grassmann collective coordinates in the neighbourhood of the bosonic solution.
Following the philosophy of moduli space dynamics, we should allow the Grassmann collective coordinates to depend on time and substitute into the Lagrangian of the theory. Performing the $x$ integral gives the effective quantum mechanical Lagrangian for the Grassmann coordinates. In this case, the relevant terms in the Lagrangian \eqref{ala} are the fermion kinetic
terms. Working to lowest order in the semi-classical expansion, the fermionic fields are, as expected, linear the Grassmann collective coordinates and the fermion kinetic terms are consequently quadratic in the coordinates. By brute force computation the effective  quantum mechanical action for the Grassmann collective coordinates is
\EQ{
S_\text{eff}&=\frac{k}{\pi}\int d^2x\, \text{STr}\,\left(-\Lambda\psi_+\dot\psi_++\Lambda\psi_-\dot\psi_-\right)\\
&=\frac{2k\sin q}{\pi}\int dt\,\left(\dot\Bu_+^*\cdot\Bu_+-\dot\Bu_-^*\cdot\Bu_-\right)\ .
}
Once again we concentrate on the $+$ sector and drop the label. 
Upon quantization, the Grassmann coordinates simply satisfy a set of anti-commutation relations:
\EQ{
\big\{\hat\Bu_i,\hat\Bu^\dagger_j\big\}=\frac{\pi}{2k\sin q}\delta_{ij}\ .
}
A Fock space can then be built by taking the $\Bu_i^\dagger$ as creation operators and $\Bu_i$ as annihilation operators. There are 4 states in the Fock space
\EQ{
|0\rangle\ ,\qquad \hat\Bu_1^\dagger|0\rangle\ ,\qquad \hat\Bu^\dagger_2|0\rangle\ ,\qquad \hat\Bu^\dagger_1\hat\Bu^\dagger_2|0\rangle\ .
}

Now once we turn on the Grassmann coordinates, the bosonic part of the soliton solution is modified even at the classical level. As we have described at the start of this section, we will consider the effects of this back-reaction to leading order because this will suffice to motivate the supersymmetry structure of the effective theory. The leading order effect of back reaction is the fact that the Grassmann odd modes carry kink, or $H_b$, charge. For example, we saw this at the perturbative level in \eqref{ncur}. We can extract the back-reaction on $q_0$ from the full soliton solution keeping terms to linear order in the bi-linear operator $\Bu^\dagger\cdot\Bu$. The effect is very simple and amounts to a shift 
\EQ{
q\to q'=q-\sin q\,\hat\Bu^\dagger\cdot\hat\Bu=q-\frac\pi{2k}\hat{\EuScript N}_f
}
in~\eqref{EffB}, where ${\EuScript N}_f$ is the fermion occupation number, which is the expectation value of 
\EQ{
\hat{\EuScript N}_f=\frac{2k\sin q}\pi \hat\Bu^\dagger\cdot\Bu\ .
}
Consequently, the bosonic and fermionic representations are correlated. Given that the bosonic occupation number is ${\EuScript N}_b=2q'k/\pi$ we have
\EQ{
{\EuScript N}\equiv{\EuScript N}_b+{\EuScript N}_f=\frac{2qk}\pi\ .
}
We present the results for the spectrum in Table \ref{states}, in particular the 
$SU(2)_f\times SU(2)_b$ representation content. The striking thing about the spectrum is that the number of bosonic and fermionic states match at $2{\EuScript N}$, and the representations are precisely the $SU(2)\times SU(2)$ content of a ``short" or 
``atypical" totally symmetric representation of the supergroup $SU(2|2)$ of dimension $2{\EuScript N}|2{\EuScript N}$ (for example, see\cite{Beisert:2005tm,Beisert:2006qh,Arutyunov:2008zt}). When the $\pm$ sectors are put together, the representation content is
\EQ{
(2{\EuScript N}|2{\EuScript N})\times (2{\EuScript N}|2{\EuScript N})
}
of $SU(2|2)^{(+)}\times SU(2|2)^{(-)}$ for ${\EuScript N}=1,2,\ldots,k$.

\TABLE[ht]{
{\begin{tabular}{ccccc}
\hline\hline
\\[-10pt]
fermionic state     &      ${\EuScript N}_f$   &  ${\EuScript N}_b$   & $SU(2)_b\times SU(2)_f$ & $\#$ states\\
\\[-10pt]
\hline
\\[-10pt]
$|0\rangle$    &   $0$    & ${\EuScript N}$ & $({\EuScript N}+1,0)$ & ${\EuScript N}+1$\\[5pt]
$\hat\Bu^\dagger_i|0\rangle$      & $1$ &      ${\EuScript N}-1$   &  $({\EuScript N},2)$  & $2{\EuScript N}$\\[5pt]
$\hat\Bu^\dagger_1\hat\Bu^\dagger_2|0\rangle$     &  $2$ &      ${\EuScript N}-2$  
& $({\EuScript N}-1,0)$ & ${\EuScript N}-1$\\[5pt]
\hline\hline
\\[-5pt]
\end{tabular}}
\label{states}
\caption{The soliton states with a given value of ${\EuScript N}$ and $q=\frac{\pi{\EuScript N}}{2k}$. The associated dimensions of the $SU(2)_b\times SU(2)_f$ representations are shown in the third column.
}}

\noindent
{\bf Manifestly supersymmetric formulation}

These results suggest that just like the original field theory, the quantum mechanical theory on the moduli has a hidden $P(SU(2|2)^{\times2})$ symmetry. In fact, we have already pointed out that the collective coordinates of the classical soliton can be thought of as the projective superspace
$\CP^{2|1}$ which can be described as the quotient
\EQ{
\CP^{2|1}\simeq \frac{SU(2|2)}{U(2|1)}\,,
}
and so has a natural group of isometries given by the left action of $SU(2|2)$. 
The collective coordinate dynamics for the $+$ sector is defined by the complete quantum mechanical action
\EQ{
S_\text{eff}=\frac{2k}\pi\int dt\,\left\{\big(q-\sin q\,\Bu^*\cdot\Bu\big)\dot\BOmega^*\cdot\BOmega+\sin q\,\dot\Bu^*\cdot\dot \Bu\right\}\ ,
\label{sdd}
}
along with the constraint $\BOmega^*\cdot\BOmega=1$. We can package the bosonic and Grassmann collective coordinates into a unit $(2|2)$ vector
\EQ{
\BZ=\left(\left[\frac{\sin q}{q}\right]^{1/2}\,\Bu\, ,\,\left[1-\frac{\sin q}{q}\Bu^*\cdot\Bu\right]^{1/2}\,\BOmega\right)
\label{sws}
}
with the constraint $\BZ^*\cdot\BZ=1$, and then \eqref{sdd} consists only of the terms of lowest power in the Grassmann odd coordinates of the effective action
\EQ{
S_\text{eff}=\frac{2kq}\pi \int dt\,\dot\BZ^*\cdot\BZ\ .
\label{swq}
}
This parameterization makes the  $SU(2|2)$ symmetry manifest. Note that it is a target space supersymmetry rather than a worldline supersymmetry. One interesting point of this is that the vector in \eqref{sws}  appears to be ``renormalized" by some specific scalar factors. Note that these ``wavefunction renormalization" factors have a well defined perturbative expansion
\EQ{
\frac{\sin q}q=1+{\cal O}(k^{-2})\,,
}
and may be a result of the schizophrenic way that we have treated the Grassmann odd and even collective coordinates. In particular, the tower of states includes the perturbative modes at the bottom since those with ${\EuScript N}_b=1$ form the fundamental representation of $SU(2|2)$. For these states the Grassmann odd and even modes are on the same footing and the renormalization factors are just $\sin q/q\sim 1$ to leading order in the perturbative expansion. 

We can quantize the system \eqref{swq} directly in a way that keeps the supersymmetry manifest.
In fact the quantization of this system is a superspace generalization of the K\"ahler quotient construction of the complex projective spaces. At the quantum level the operators $\hat\BZ$ satisfy a set or (anti-)commutation relations
\EQ{
[\hat\BZ_i,\hat\BZ_j^\dagger]_\pm=\frac{\pi}{2kq}\delta_{ij}\ .
}
The Hilbert space consists of states of the form
\EQ{
\hat\BZ_{i_1}^\dagger\hat\BZ_{i_2}^\dagger\cdots\hat\BZ_{i_{\EuScript N}}^\dagger|0\rangle\ ,
\label{fhh}
}
with $q$ quantized as
\EQ{
q=\frac{\pi{\EuScript N}}{2k}\ .
}
The states \eqref{fhh} transform as the atypical symmetric representation of $SU(2|2)$ of dimension $2{\EuScript N}|2{\EuScript N}$.

\acknowledgments

We would like to thank David Schmidtt for useful discussions, and Arkady Tseytlin and Ben Hoare for useful discussions and for letting us see a draft of their new paper on the $S$-matrix of these theories~\cite{Hoare:2011fj}.

\noindent
JLM acknowledges the support of MICINN (FPA2008-01838 and 
FPA2008-01177), Xunta de Galicia (Consejer\'\i a de Educaci\'on and INCITE09.296.035PR), the 
Spanish Consolider-Ingenio 2010
Programme CPAN (CSD2007-00042), and FEDER.

\noindent  
TJH would like to acknowledge the support of STFC grant
ST/G000506/1.

\startappendix

\Appendix{Supersymmetry Off-Shell}

In this appendix we show that the supersymmetry transformations \eqref{susv} and \eqref{eqd} that we derived as on-shell symmetries of the equations-of-motion are also off-shell symmetries of the gauge-fixed action. The complication is that the transformations are non-local and this is probably why the supersymmetry in the quantum theory becomes $q$ deformed and so is ultimately not a true symmetry.

If we fix the gauge, as in \cite{Hoare:2009rq,Hoare:2009fs,Hoare:2010fb}, by taking $A_+=0$, then $A_-$ remains as a Lagrange multiplier for the constraint
\EQ{
\big(\gamma^{-1}\partial_+\gamma\big)^\perp+2\Lambda\psi_+\psi_+=0\ .
\label{consa}
}
The discussion in Section 3, where we constructed the conserved supercurrents in the on-shell gauge $A_\mu=0$, requires some modification when $A_-\neq0$. Essentially, it amounts to including $A_-$ in $h_-(z)$:
\EQ{
h_-(z)=A_-+\sum_{s<0}h_{s,-}z^s\ .
}
The zero-curvature condition \eqref{Zero2} still applies, but now it implies that the supercurrent is only covariantly conserved $D^\mu{\cal G}^{(+)}_\mu=0$, where the covariant derivative includes the component $A_-$:
\EQ{
\Phi^{-1}[{\cal L}_+,{\cal L}_-]\Phi&=
\bigl[\partial_+ -z^2\Lambda+ h_+(z),  \partial_-  + h_-(z)\bigr]\\
&=\partial_+A_-+z^{-1}(D_-{\cal G}_+^{(+)}+\partial_+{\cal G}_-^{(+)})-[z^2\Lambda,h_-(z)]+{\cal O}(z^{-2})\ .
\label{jss}
}
Note, that $h_+(z)\in\hat{\mathfrak f}^\perp$ off-shell, but $h_-(z)$ is only in $\hat{\mathfrak f}^\perp$ on-shell which explains the appearance of the commutator term.

The issue before us is to show that the supersymmetry is a Noether symmetry of the theory and consequently a symmetry---albeit a non-local one---of the action. For a general theory, the variation of the action is, schematically,
\EQ{
\delta  S&= \int d^2x\Big[\frac{\partial\LAG}{\partial\varphi} \delta  \varphi + \frac{\partial \LAG}{\partial(\partial_\mu\varphi)}\partial_\mu\delta  \varphi\Big]\\
 &=\int d^2x\Big[\frac{\delta\LAG}{\delta\varphi}\,\delta\varphi+\partial_\mu\left(\frac{\partial  \LAG}{\partial(\partial_\mu\varphi)}\delta  \varphi\right)\Big]\ .
}
where
\EQ{
\frac{\delta\LAG}{\delta\varphi}\equiv-\partial_\mu\frac{\partial\LAG}{\partial\partial_\mu\varphi}+\frac{\partial\LAG}{\partial\varphi}\approx0
}
are the equations-of-motion. So if
\EQ{
\frac{\delta\LAG}{\delta\varphi}\,\delta\varphi=\partial_\mu R^\mu\ ,
}
for a particular variation $\delta\varphi$,
then the variation of the Lagrangian is a total derivative and, assuming suitable behaviour at infinity, the action is invariant under the symmetry $\delta\varphi$.

In the present context, assuming that $\delta A_-=0$, we have
\EQ{
\frac{\delta\LAG}{\delta\varphi}\,\delta\varphi=\Big\langle[{\cal L}_+(z),{\cal L}_-(z)],\delta B(z)\Big\rangle\ ,
\label{xpp}
}
where
\EQ{
\delta B(z)=z^{-1}[\delta\psi_+,\Lambda]+\gamma^{-1}\delta\gamma
+z\gamma^{-1}[\delta\psi_-,\Lambda]\gamma
}
and we have defined the inner product on the affine algebra
\EQ{
\Big\langle A(z),B(z)\Big\rangle=\text{STr}\,\Big(\sum_nA_nB_{-n}\Big)\ .
}
Using the zero-curvature condition \eqref{Zero2}, or~\eqref{jss}, this becomes
\EQ{
\frac{\delta\LAG}{\delta\varphi}\,\delta\varphi=\Big\langle[\partial_+-z^2\Lambda+h_+(z),\partial_-+h_-(z)],\Phi(z)^{-1}\delta B(z)\Phi(z)\Big\rangle\ .
}
Now if
\EQ{
\Phi(z)\delta B(z)\Phi^{-1}=z\varepsilon_++q^\perp+{\cal O}(z^{-2})
\label{gew}
}
for  $\varepsilon_+\in\mathfrak f_1$ and $q^\perp\in\mathfrak h$,
then, using \eqref{jss}, we have, assuming that $\varepsilon_+$ is constant,
\EQ{
\frac{\delta\LAG}{\delta\varphi}\,\delta\varphi&=\text{STr}\,\big(
q^\perp\partial_+A_-+\varepsilon_+(D_-{\cal G}^{(+)}_++\partial_+{\cal G}^{(+)}_-)\big)\\
&=\partial_+\,\text{STr}\,(\varepsilon_+{\cal G}^{(+)}_-+A_-q^\perp)+
\partial_-\,\text{STr}\,(\varepsilon_+{\cal G}^{(+)}_+)
}
where $q^\perp$ is given by the non-local expression
\EQ{
\partial_+q^\perp=-[\varepsilon_+,{\cal G}^{(+)}_+]\ .
}
Notice that the terms in \eqref{jss} involving the commutator with $\Lambda$ do not contribute because $\varepsilon_+$ and $q^\perp$ are in $\hat{\mathfrak f}^\perp$.
So the variation of the Lagrangian is a total derivative and the action is invariant.
Note that in the on-shell gauge $A_\mu=0$, the above implies $\partial_-q^\perp=[\varepsilon_+,{\cal G}^{(+)}_-]$ which is \eqref{eqd}.
We can then solve \eqref{gew} for the variations of the fields to reproduce \eqref{susv}.
So this proves that the action is invariant under the non-local supersymmetry transformations.

The transformations associated to the current ${\cal G}^{(-)}_\mu$ follow in a similar way from 
\EQ{
\tilde\Phi(z)\delta B(z)\tilde\Phi^{-1}=z^{-1}\varepsilon_-+\tilde q^\perp+{\cal O}(z^{2})
\label{gew2}
}
with
\EQ{
\partial_-\tilde q^\perp=-[\varepsilon_-,{\cal G}^{(-)}_+]\ .
}

\end{document}